# A Taxonomy and Survey of Energy-Efficient Data Centers and Cloud Computing Systems

Anton Beloglazov[1], Rajkumar Buyya[1], Young Choon Lee[2], and Albert Zomaya[2]

[1] Cloud Computing and Distributed Systems (CLOUDS) Laboratory
Department of Computer Science and Software Engineering
The University of Melbourne, Australia

[2] Centre for Distributed and High Performance Computing
School of Information Technologies
The University of Sydney, Australia

## Abstract

Traditionally, the development of computing systems has been focused on performance improvements driven by the demand of applications from consumer, scientific and business domains. However, the ever increasing energy consumption of computing systems has started to limit further performance growth due to overwhelming electricity bills and carbon dioxide footprints. Therefore, the goal of the computer system design has been shifted to power and energy efficiency. To identify open challenges in the area and facilitate future advancements it is essential to synthesize and classify the research on power and energy-efficient design conducted to date. In this work we discuss causes and problems of high power / energy consumption, and present a taxonomy of energy-efficient design of computing systems covering the hardware, operating system, virtualization and data center levels. We survey various key works in the area and map them to our taxonomy to guide future design and development efforts. This chapter is concluded with a discussion of advancements identified in energy-efficient computing and our vision on future research directions.

# Table of Contents

# 1 Introduction

The primary focus of designers of computing systems and the industry has been on the improvement of the system performance. According to this objective the performance has been steadily growing driven by more efficient system design and increasing density of the components described by Moore's law [1]. Although the performance per watt ratio has been constantly rising, the total power draw by computing systems is hardly decreasing. Oppositely, it has been increasing every year that can be illustrated by the estimated average power use across three classes of servers presented in Table 1 [2]. If this trend continues, the cost of the energy consumed by a server during its lifetime will exceed the hardware cost [3]. The problem is even worse for large-scale compute infrastructures, such as clusters and data centers. It was estimated that in 2006 IT infrastructures in the US consumed about 61 billion kWh for the total electricity cost about 4.5 billion dollars [4]. The estimated energy consumption is more than double from what was consumed by IT in 2000. Moreover, under current efficiency trends the energy consumption tends to double again by 2011, resulting in 7.4 billion dollars annually.

Table 1. Estimated average power consumption per server class (W/Unit) from 2000 to 2006 [2].

| Server class | 2000 | 2001 | 2002 | 2003 | 2004 | 2005 | 2006 |
|---|---|---|---|---|---|---|---|
| Volume | 186 | 193 | 200 | 207 | 213 | 219 | 225 |
| Mid-range | 424 | 457 | 491 | 524 | 574 | 625 | 675 |
| High-end | 5,534 | 5,832 | 6,130 | 6,428 | 6,973 | 7,651 | 8,163 |

The energy consumption is not only determined by the efficiency of the physical resources, but it is also dependent on the resource management system deployed in the infrastructure and efficiency of applications running in the system. This interconnection of the energy consumption and different levels of computing systems can be seen from Figure 1. Energy efficiency impacts end users in terms of resource usage costs, which are typically determined by the Total Cost of Ownership (TCO) incurred by a resource provider. Higher power consumption results not only in boosted electricity bills, but also in additional requirements to a cooling system and power delivery infrastructure, i.e. Uninterruptible Power Supplies (UPS), Power Distribution Units (PDU), etc. With the growth of computer components density, the cooling problem becomes crucial, as more heat has to be dissipated for a square meter. The problem is especially important for 1U and blade servers. These form factors are the most difficult to cool because of high density of the components, and thus lack of space for the air flow. Blade servers give the advantage of more computational power in less rack space. For example, 60 blade servers can be installed into a standard 42U rack [5]. However, such system requires more than 4,000 W to supply the resources and cooling system compared to the same rack filled by 1U servers consuming 2,500 W. Moreover, the peak power consumption tends to limit further performance improvements due to constraints of power distribution facilities. For example, to power a server rack in a typical data center, it is necessary to provide about 60 Amps [6]. Even if the cooling problem can be addressed for the future systems, it is likely that delivering current in such data centers will reach the power delivery limits.

Apart from the overwhelming operating costs and the Total Cost of Acquisition (TCA), another rising concern is the environmental impact in terms of carbon dioxide ($CO_2$) emissions caused by high energy consumption. Therefore, the reduction of power and energy consumption has become a first-order objective in the design of modern computing systems. The roots of energy-efficient computing, or Green IT, practices can be traced back to 1992, when the U.S. environmental protection Agency launched Energy Star, a voluntary labelling program which is designed to identify and promote energy-efficient products in order to reduce the greenhouse gas emissions. Computers and monitors were the first labelled products. This has led to the widespread adoption of the sleep mode in electronic devices. At that time the term "green computing" was introduced to refer to energy-efficient personal computers [7]. At the same time, the Swedish confederation of professional employees has developed the TCO certification program – a series of end user and environmental requirements for IT equipment including video adapters, monitors, keyboards, computers, peripherals, IT systems and even mobile phones. Later, this program has been extended to include requirements on ergonomics, magnetic and electrical field emission levels, energy consumption, noise level and use of hazardous compounds in hardware. The Energy Star program was revised in October 2006 to include stricter efficiency requirements for computer equipment, along with a tiered ranking system for approved products.

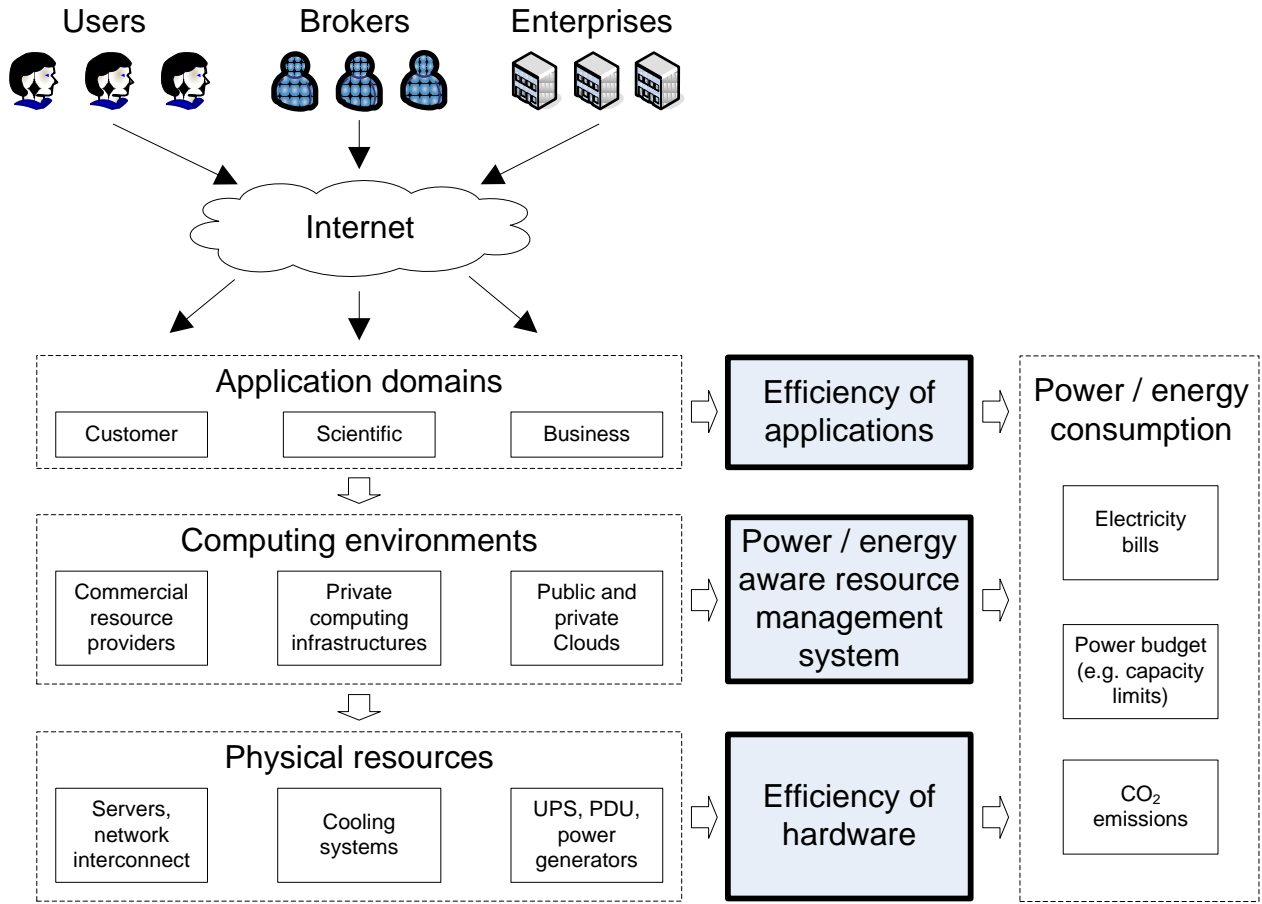

Figure 1. Energy consumption at different levels in computing systems.

There are a number of industry initiatives aiming at the development of standardized methods and techniques for reduction of the energy consumption in computer environments. They include Climate Savers Computing Initiative (CSCI), Green Computing Impact Organization, Inc. (GCIO), Green Electronics Council, The Green Grid, International Professional Practice Partnership (IP3), with membership of companies such as AMD, Dell, HP, IBM, Intel, Microsoft, Sun Microsystems and VMware.

Energy-efficient resource management has been first introduced in the context of battery feed mobile devices, where energy consumption has to be reduced in order to improve the battery lifetime. Although techniques developed for mobile devices can be applied or adapted for servers and data centers, this kind of systems requires specific methods. In this chapter we will discuss ways to reduce power and energy consumption in computing systems, as well as recent research works that deal with power and energy efficiency at the hardware and firmware, Operating System (OS), virtualization, and data center levels. The main objective of this work is to give an overview of the recent research advancements in energy-efficient computing, identify common characteristics and classify the approaches. On the other hand, the aim is to show the level of development in the area and discuss open research challenges and direction for future work. The reminder of this chapter is organized as follows: in the next Section power and energy models are introduced; in Section 3 we discuss problems caused by high power and energy consumption; in Sections 4-8 we present the taxonomy and survey of the research in energy-efficient design of computing systems, followed by a conclusion and future work directions in Section 9.

# 2 Power and Energy Models

To understand power and energy management mechanisms it is essential to clearly distinguish the background terms. Electric current is the flow of electric charge measured in Amperes (Amps). Amperes define the amount of electric charge transferred by a circuit per second. Power and energy can be defined in terms of work that a system performs. Power is the rate at which the system performs the work, while energy is the total amount of work performed over a period of time. Power and energy are measured in watts (W) and watt-hour (Wh) respectively. Work is done at the rate of one watt when one Ampere is transferred through a potential difference of one volt. A kilowatt-hour (kWh) is the amount of energy equivalent to a power of 1 kilowatt (1000 watts) running for 1 hour. Formally, power and energy can be defined as in (1) and (2).

$$P = \frac{W}{T}, \tag{1}$$

$$E = P \cdot T, \tag{2}$$

where $P$ is power, $T$ is a period of time, $W$ is the total work performed in that period of time, and $E$ is energy. The difference between power and energy is very important, because reduction of the power consumption does not always reduce the consumed energy. For example, the power consumption can be decreased by lowering the CPU performance. However, in this case a program may require longer time to complete its execution consuming the same amount of energy. On one hand, reduction of the peak power consumption will result in decreased costs of the infrastructure provisioning, such as costs associated with capacities of UPS, PDU, power generators, cooling system, and power distribution equipment. On the other hand, decreased energy consumption will lead to reduction of the electricity bills. The energy consumption can be reduced temporarily using Dynamic Power Management (DPM) techniques or permanently applying Static Power Management (SPM). DPM utilizes knowledge of the real-time resource usage and application workloads to optimize the energy consumption. However, it does not necessarily decrease the peak power consumption. In contrast, SPM includes the usage of highly efficient hardware equipment, such as CPUs, disk storage, network devices, UPS and power supplies. These structural changes usually reduce both the energy and peak power consumption.

## 2.1 Static and Dynamic Power Consumption

The main power consumption in Complementary Metal-Oxide-Semiconductor (CMOS) circuits comprises static and dynamic power. The static power consumption, or leakage power, is caused by leakage currents that are present in any active circuit, independently of clock rates and usage scenarios. This static power is mainly determined by the type of transistors and process technology. Reduction of the static power requires improvement of the low-level system design; therefore, it is not in the focus of this chapter. More details about possible ways to improve the energy efficiency at this level can be found in the survey by Venkatachalam and Franz [8].

Dynamic power consumption is created by circuit activity (i.e. transistor switches, changes of values in registers, etc.) and depends mainly on a specific usage scenario, clock rates, and I/O activity. The sources of the dynamic power consumption are short-circuit current and switched capacitance. Short-circuit current causes only 10-15% of the total power consumption and so far no way has been found to reduce this value without compromising the performance. Switched capacitance is the primary source of the dynamic power consumption; therefore, the dynamic power consumption can be defined as in (3).

$$P_{dynamic} = a \cdot C \cdot V^2 \cdot f,  \qquad (3)$$

where $a$ is the switching activity, $C$ is the physical capacitance, $V$ is the supply voltage, and $f$ is the clock frequency. The values of switching activity and capacitance are determined by the low-level system design. Whereas combined reduction of the supply voltage and clock frequency lies in the roots of the widely adopted DPM technique called Dynamic Voltage and Frequency Scaling (DVFS). The main idea of this technique is to intentionally down-scale CPU performance, when it is not fully utilized, by decreasing the voltage and frequency of the CPU that in ideal case should result in cubic reduction of the dynamic power consumption. DVFS is supported by most modern CPUs including mobile, desktop and server systems. We will discuss this technique in detail in Section 5.2.1.

## 2.2 Sources of Power Consumption

According to data provided by Intel Labs [5] the main part of power consumed by a server is drawn by the CPU, followed by the memory and losses due to the power supply inefficiency (Figure 2). The data show that the CPU no longer dominates power consumption by a server. This resulted from continuous improvement of the CPU power efficiency and application of power saving techniques (e.g. DVFS) that enable active low-power modes. In these modes a CPU consumes a fraction of the total power, while preserving the ability to execute programs. As a result, current desktop and server CPUs can consume less than 30% of their peak power at low-activity modes leading to dynamic power range of more than 70% of the peak power [9]. In contrast, dynamic power ranges of all other server's components are much narrower: less than 50% for DRAM, 25% for disk drives, 15% for network switches, and negligible for other components [10]. The reason is that only the CPU supports active low-power modes, whereas other components can only be completely or partially switched off. However, the performance overhead of transition between active and inactive modes is substantial. For example, a disk drive in a spun-down, deep-sleep mode consumes almost no power, but a transition to active mode incurs a latency that is 1,000 times higher than regular access latency. Power inefficiency of the server's components in the idle state leads to a narrow overall dynamic power range of 30%. This means that even if a server is completely idle, it will still consume more than 70% of its peak power.

Another reason for reduction of the fraction of power consumed by the CPU relatively to the whole system is adoption of multi-core architectures. Multi-core processors are much more efficient than conventional. For example, servers built with recent Quad-core Intel Xeon processor can deliver 1.8 teraflops at peak performance, using less than 10 kW of power. To compare with, Pentium processors in 1998 would consume about 800 kW to achieve the same performance [5].

Adoption of multi-core CPUs along with the increasing use of virtualization technologies and data-intensive applications resulted in growing amount of memory in servers. In contrast to the CPU, DRAM has narrower dynamic power range and power consumption by memory chips is increasing. Memory is packaged in dual in-line memory modules (DIMMs), and power consumption by these modules varies from 5 W to 21 W per DIMM, for DDR3 and Fully Buffered DIMM (FB-DIMM) memory technologies [5]. Power consumption by a server with eight 1 GB DIMMs is about 80 W. Modern large servers currently use 32 or 64 DIMMs that leads to power consumption by memory higher than by CPUs. Most of the power management techniques are focused on the CPU; however, constantly increasing frequency and capacity of memory chips raise the cooling requirements apart from the problem of high energy consumption. These facts make memory one of the most important server components that have to be efficiently managed. New techniques and approaches to the reduction of the memory power consumption have to be developed in order to address this problem.

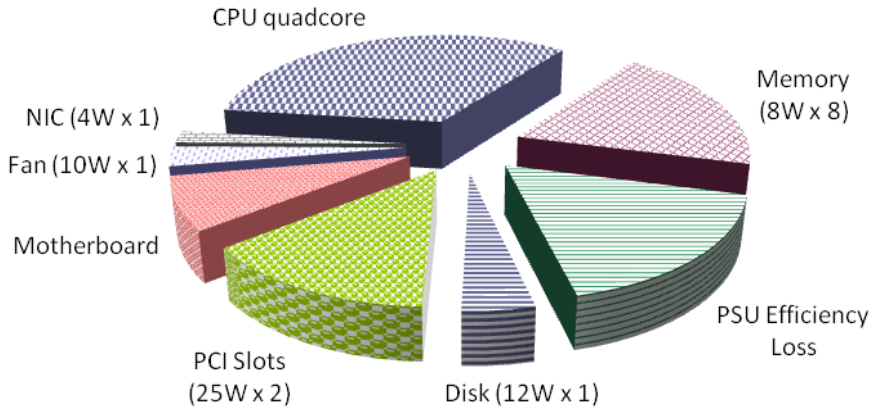

Figure 2. Power consumption by server's components [5].

Power supplies transform alternating current (AC) into direct current (DC) to feed server's components. This transformation leads to significant power losses due to the inefficiency of the current technology. The efficiency of power supplies depends on their load. They achieve the highest efficiency at loads within the range of 50-75%. However, most data centers create a load of 10-15% wasting the majority of the consumed electricity and leading to average power losses of 60-80% [5]. As a result, power supplies consume at least 2% of all U.S. electricity production. More efficient power supply design can save more than a half of the energy consumption.

The problem of low average utilization applies to disk storages, especially when disks are attached to servers in a data center. However, this can be addressed by moving the disks to an external centralized storage array. Nevertheless, intelligent policies have to be used to efficiently manage a storage system containing thousands of disks. This creates another direction for the research work aimed at optimization of resource, power and energy usage in server farms and data centers.

## 2.3 Modeling Power Consumption

To develop new policies for DPM and understand their impact, it is necessary to create a model of dynamic power consumption. Such a model has to be able to predict the actual value of the power consumption based on some run-time system characteristics. One of the ways to accomplish this is to utilize power monitoring capabilities that are built-in modern computer servers. This instrument provides the ability to monitor power usage of a server in real time and collect accurate statistics about the power usage. Based on this data it is possible to derive a power consumption model for a particular system. However, this approach is complex and requires collection of the statistical data for each target system.

Fan et al. [10] have found a strong relationship between the CPU utilization and total power consumption by a server. The idea behind the proposed model is that the power consumption by a server grows linearly with the growth of CPU utilization from the value of power consumption in the idle state up to the power consumed when the server is fully utilized. This relationship can be expressed as in (4).

$$P(u) = P_{idle} + (P_{busy} - P_{idle}) * u, \qquad (4)$$

where $P$ is the estimated power consumption, $P_{idle}$ is the power consumption by an idle server, $P_{busy}$ is the power consumed by the server when it is fully utilized, and $u$ is current CPU utilization. The authors have also proposed an empirical non-linear model given in (5).

$$P(u) = P_{idle} + (P_{busy} - P_{idle}) \cdot (2u - u^r), \quad (5)$$

where $r$ is a calibration parameter that minimizes the square error and has to be obtained experimentally. For each class of machines of interest a set of calibration experiments must be performed to fine tune the model.

Extensive experiments on several thousands of nodes under different types of workloads (Figure 3) have shown that the derived models accurately predict the power consumption by server systems with the error below 5% for the linear model and 1% for the empirical model. The calibration parameter $r$ has been set to 1.4 for the presented results. These precise results can be explained by the fact that CPU is the main power consumer in servers and, in contrast to CPU, other system components have narrow dynamic power ranges or their activities correlate with the CPU activity (e.g. I/O, memory). For example, current server processors can reduce power consumption up to 70% by switching to low power-performance modes [9]. However, dynamic power ranges of other components are much narrower: less than 50% for DRAM, 25% for disk drives, and 15% for network switches.

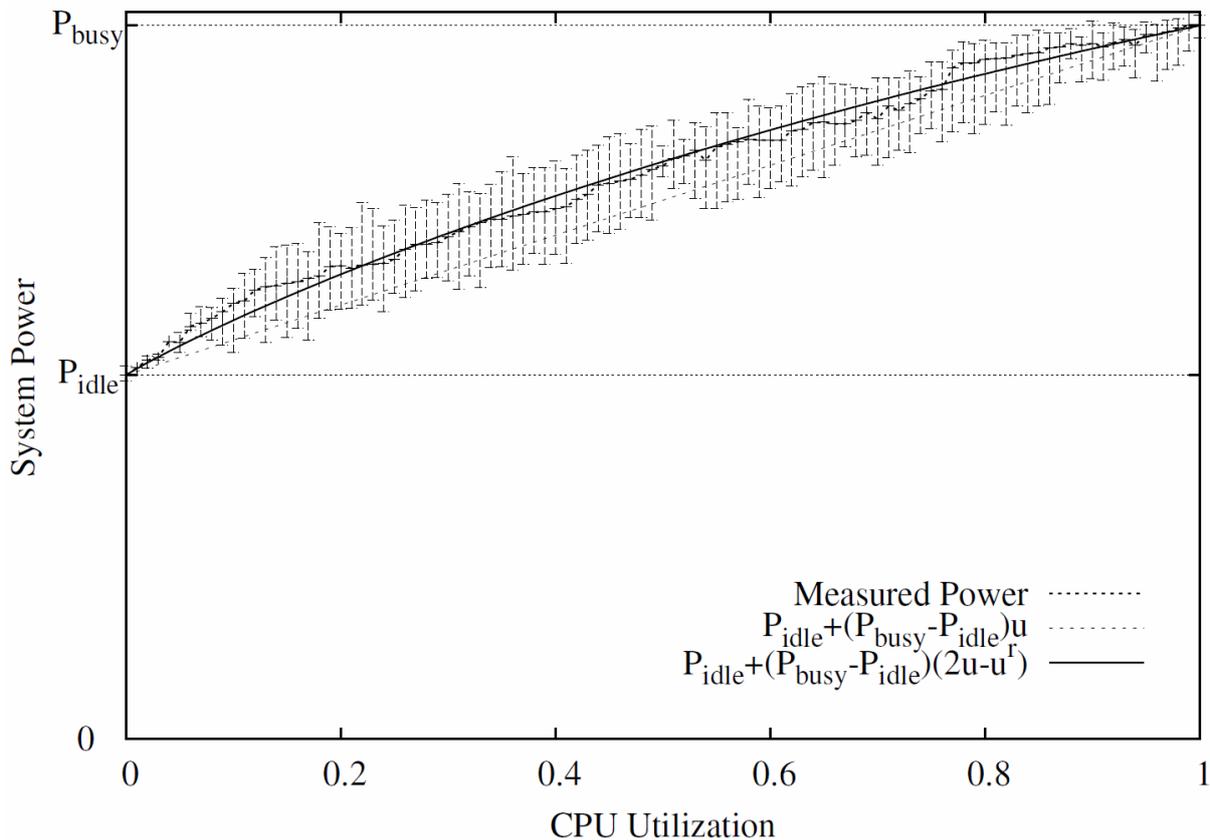

Figure 3. Power consumption to CPU utilization relationship [10].

This accurate and simple power model enables easy prediction of the power consumption by a server supplied with CPU utilization data and power consumption values at idle and maximum CPU utilization states. Therefore, it is especially important that the increasing number of server manufactures publish actual power consumption figures for their systems at different utilization levels [11]. This is driven by the adoption of the ASHRAE Thermal Guideline [12] that recommends providing power ratings for minimum, typical and full utilization.

Dhiman et al. [13] have found that although regression models based on just CPU utilization are able to provide reasonable prediction accuracy for CPU-intensive workloads, they tend to be considerably inaccurate for prediction of power consumption caused by I/O- and memory-intensive applications. The authors have proposed a power modeling methodology based on Gaussian Mixture Models that predicts power consumption by a physical machine running multiple VM instances. To perform predictions, in addition to CPU utilization the model relies on run-time workload characteristics, such as the number of Instructions Per Cycle (IPC) and the number of Memory accesses Per Cycle (MPC). The proposed approach requires a training phase to perceive the relationship between the metrics of the workload and the power consumption. The authors have evaluated the proposed model via experimental studies involving different types of the workload. The obtained experimental results have shown that the model predicts the power consumption with high accuracy (<10% prediction error), which is consistent over all the tested workloads. The proposed model outperforms regression models by a factor of 5 for particular types of the workload, which proves the importance of architectural metrics like IPC and MPC as compliments to CPU utilization for prediction of the power consumption.

# 3 Problems of High Power and Energy Consumption

The energy consumption by computing facilities rises various monetary, environmental and system performance concerns. A recent study on the power consumption of server farms [2] shows that in 2005 the electricity use by servers worldwide – including their associated cooling and auxiliary equipment – costed US$7.2bn. The study also indicates that the electricity consumption in that year had doubled as compared with consumption in 2000. Clearly, there are environmental issues with the generation of electricity. The number of transistors integrated into today's Intel Itanium 2 processor reaches to nearly 1 billion. If this rate continues, the heat (per square centimetre) produced by future processors would exceed that of the surface of the Sun [14], resulting in poor system performance. The scope of energy-efficient design is not limited to main computing components (e.g., processors, storage devices and visualization facilities), but it can expand into a much larger range of resources associated with computing facilities including auxiliary equipments, water used for cooling and even physical/floor space that these resources occupy.

While recent advances in hardware technologies including low-power processors, solid state drives and energy-efficient monitors have alleviated the energy consumption issue to a certain degree, a series of software approaches have significantly contributed to the improvement of energy efficiency. These two approaches (hardware and software) should be seen as complementary rather than competitive. User awareness is another non-negligible factor that should be taken into account when discussing Green IT. User awareness and behavior in general considerably affect computing workload and resource usage patterns; this in turn has a direct relationship with the energy consumption of not only core computing resources, but also auxiliary equipment, such as cooling/air conditioning systems. For example, a computer program developed without paying much attention to its energy efficiency may lead to excessive energy consumption and it may contribute to more heat emission resulting in increases in the energy consumption for cooling.

Traditionally, power and energy-efficient resource management techniques have been applied to mobile devices. It was dictated by the fact that such devices are usually battery-powered and it is essential to consider power and energy management to improve their lifetime. However, due to continuous growth of power and energy consumption by servers and data centers, the focus of power and energy management techniques has been switched to these systems. Even though the problems caused by high power and energy consumption are interconnected, they have their specifics and have to be considered separately. The difference is that the peak power consumption determines the cost of the infrastructure required to maintain the system's operation, whereas the energy consumption accounts for electricity bills. Let us discuss each of these problems in detail.

## 3.1 High Power Consumption

The main reason of the power inefficiency in data centers is low average utilization of the resources. We have used data provided as a part of the CoMon project[1], a monitoring infrastructure for PlanetLab[2]. We have used the data of the CPU utilization by more than a thousand servers located at more than 500 places around the world. The data have been collected each five minutes during the period from the 10th to 19th of May 2010. The distribution of the data over the mentioned 10 days along with the characteristics of the distribution are presented in Figure 4. The data confirm the observation made by Barroso and Holzle [9]: the average CPU utilization is below 50%. The mean value of the CPU utilization is 36.44% with 95% confidence interval: (36.40%, 36.47%). The main run-time reasons of underutilization in data centers are variability of the workload and statistical effects. Modern service applications cannot be kept on fully utilized servers, as even non-significant workload fluctuation will lead to performance degradation and failing to provide the expected QoS. On the other hand, servers in a non-virtualized data center are unlikely to be completely idle because of background tasks (e.g. incremental backups), or distributed data bases or file systems. Data distribution helps to tackle load-balancing problem and improves fault-tolerance. Furthermore, despite the fact that the resources have to be provisioned to handle theoretical peak loads, it is very unlikely that all the servers of a large-scale data centers will be fully utilized simultaneously.

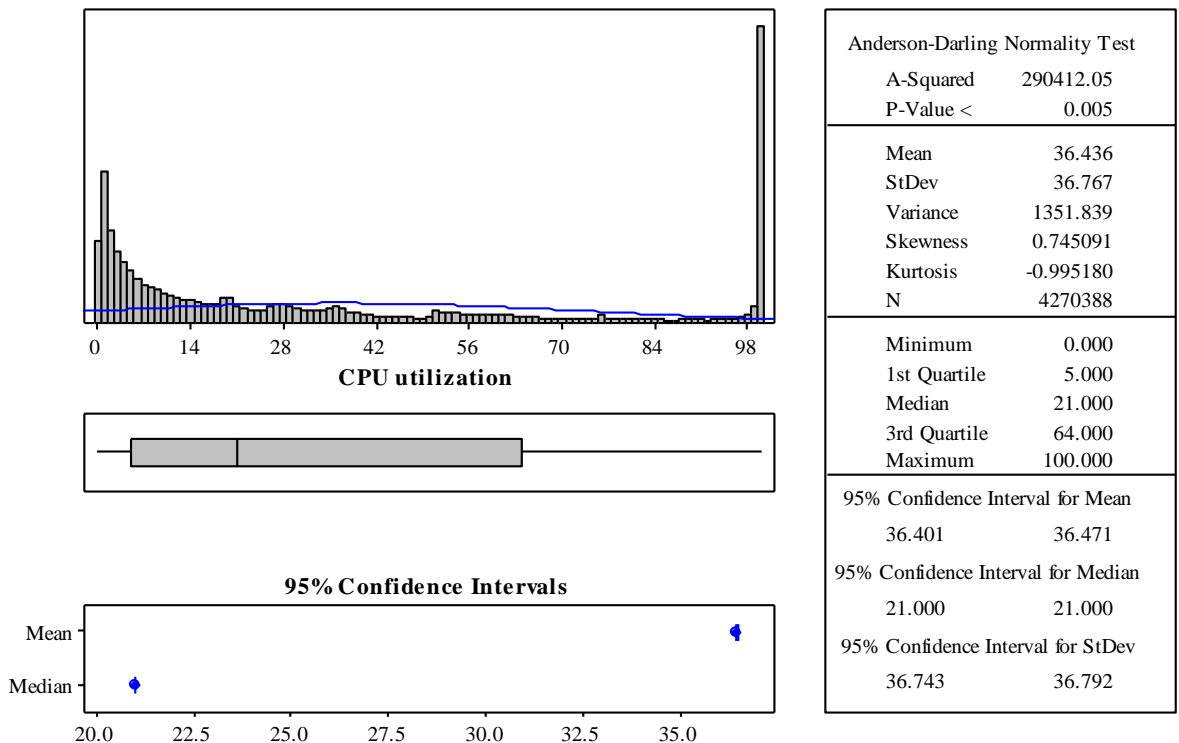

Figure 4. The CPU utilization of more than 1000 PlanetLab nodes over a period of 10 days.

Systems where average utilization of resources less than 50% represent huge inefficiency, as most of the time only a half of the resources are actually in use. Although the resources on average are utilized by less than 50%, the infrastructure has to be built to handle the peak load, which rarely occurs in practice. In such systems the cost of over-provisioned capacity is very significant and includes expenses on additional capacity of the cooling system, PDU, generators, power delivery facilities, UPS, etc. The less average resource utilization in a data center, the more expensive it becomes as a part of the Total Cost of Ownership (TCO), as it has to support peak loads and meet the requirements to the peak power consumption. For example, if a data center operates at 85% of its peak capacity on average, the cost of building the data center (in terms of the building cost per Watt of the average power consumption) will still exceed the electricity cost for ten years of operation [10]. Moreover, peak power consumption can constrain further growth of power density, as power requirements already reach 60 Amps for a server rack [6]. If this tendency continues, further performance improvements can be bounded by the power delivery capabilities.

---
[1] http://comon.cs.princeton.edu/
[2] http://www.planet-lab.org/

Another problem of high power consumption and increasing density of server's components (i.e. 1U, blade servers) is the heat dissipation. Much of the electrical power consumed by the computing resources gets turned into heat. The amount of heat produced by an integrated circuit depends on how efficient the component's design is, and the voltage and frequency at which the component operates. The heat generated by the resources has to be dissipated to keep them within their safe thermal state. Overheating of the components can lead to decreased lifetime and high error-proneness. Moreover, power is required to feed the cooling system operation. For each watt of power consumed by computing resources an additional 0.5 to 1 W is required for the cooling system [6]. For example, to dissipate 1 W consumed by a High-Performance Computing (HPC) system at the Lawrence Livermore National Laboratoy (LLNL), 0.7 W of additional power is needed for the cooling system [15]. This fact justifies the significant concern about efficiency and real-time adaptation of cooling system operation. Moreover, modern high density servers, such as 1U and blade servers, further complicate cooling because of the lack of space for airflow within the packages.

## 3.2 High Energy Consumption

Considering the power consumption, the main problem is the minimization of the peak power required to feed a completely utilized system. In contrast, the energy consumption is defined by the average power consumption over a period of time. Therefore, the actual energy consumption by a data center does not affect the cost of the infrastructure. On the other hand, it is reflected in the electricity cost consumed by the system during the period of operation, which is the main component of a data center's operating costs. Furthermore, in most data centers 50% of consumed energy never reaches the computing resources: it is consumed by the cooling facilities or dissipated in conversions within the UPS and PDU systems. With the current tendency of continuously growing energy consumption and costs associated with it, the point when operating costs exceed the cost of computing resources themselves in few years can be reached soon. Therefore, it is crucial to develop and apply energy-efficient resource management strategies in data centers.

Except for high operating costs, another problem caused by growing energy consumption is high carbon dioxide ($CO_2$) emissions, which contribute to the global warming. According to Gartner [16] in 2007 the Information and Communications Technology (ICT) industry was responsible for about 2% of global $CO_2$ emissions that is equivalent to the aviation. According to the estimation by the U.S. Environmental Protection Agency (EPA), current efficiency trends lead to the increase of annual $CO_2$ emissions from 42.8 million metric tons (MMTCO2) in 2007 to 67.9 MMTCO2 in 2011. Intense media coverage has raised the awareness of people around the climate change and greenhouse effect. More and more customers start to consider the "green" aspect in selecting products and services. Besides the environmental concern, businesses have begun to face risks caused by being non-environmentally friendly. Reduction of $CO_2$ footprints is an important problem that has to be addressed in order to facilitate further advancements in computing systems.

# 4 Taxonomy of Power / Energy Management in Computing Systems

Large volume of research work has been done in the area of power and energy-efficient resource management in computing systems. As power and energy management techniques are closely connected, from this point we will refer to them as power management. As shown in Figure 5, from the high level power management techniques can be divided into static and dynamic. From the hardware point of view, Static Power Management (SPM) contains all the optimization methods that are applied at the design time at the circuit, logic, architectural and system levels [17]. Circuit level optimizations are focused on the reduction of switching activity power of individual logic-gates and transistor level combinational circuits by the application of a complex gate design and transistor sizing. Optimizations at the logic level are aimed at the switching activity power of logic-level combinational and sequential circuits. Architecture level methods include the analysis of the system design and subsequent incorporation of power optimization techniques in it. In other words, this kind of optimization refers to the process of efficient mapping of a high-level problem specification onto a register-transfer level design. Apart from the optimization of the hardware-level system design, it is extremely important carefully consider the implementation of programs that are supposed to run in the system. Even with perfectly designed hardware, poor software design can lead to dramatic performance and power losses. However, it is impractical or impossible to analyse power consumption caused by large programs at the operator level, as not only the process of compilation or code-generation but also the order of instructions can have an impact on power consumption. Therefore, indirect estimation methods can be applied. For example, it has been shown that faster code almost always implies lower energy consumption [18]. Nevertheless, methods for guaranteed synthesizing of optimal algorithms are not available, and this is a very difficult problem.

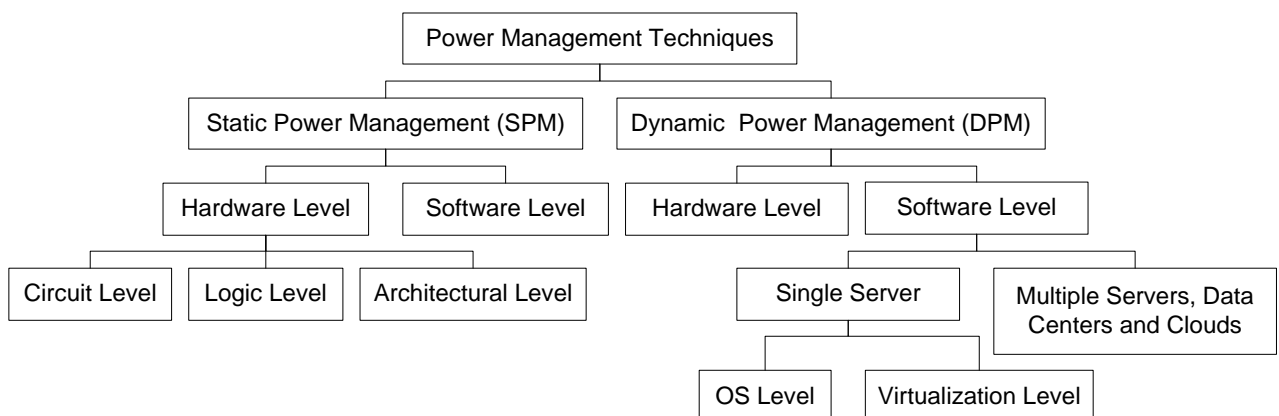

Figure 5. High level taxonomy of power and energy management.

This chapter focuses on DPM techniques that include methods and strategies for run-time adaptation of a system's behavior according to current resource requirements or any other dynamic characteristic of the system's state. The major assumption enabling DPM is that systems experience variable workloads during the operation time allowing the dynamic adjustment of power states according to current performance requirements. The second assumption is that the workload can be predicted to a certain degree. As shown on Figure 5, DPM techniques can be distinguished by the level at which they are applied: hardware or software. Hardware DPM varies for different hardware components, but usually can be classified as Dynamic Performance Scaling (DPS), such as DVFS, and partial or complete Dynamic Component Deactivation (DCD) during periods of inactivity. In contrast, software DPM techniques utilize interface to the system's power management and according to their policies apply hardware DPM. The introduction of the Advanced Power

Management (APM)[3] and its successor, the Advanced Configuration and Power Interface (ACPI)[4], have drastically simplified the software power management and resulted in broad research studies in this area. The problem of power efficient resource management has been investigated in different contexts of device specific management, OS level management of virtualized and non-virtualized servers, followed by multiple-node system, such as homogeneous and heterogeneous clusters, data centers and Clouds.

DVFS creates a broad dynamic power range for the CPU enabling extremely low-power active modes. This flexibility has lead to the wide adoption of this technique and appearance of many policies that scale CPU performance according to current requirements, while trying to minimize performance degradation [19]. Subsequently, these techniques have been extrapolated on multiple-server systems providing coordinated performance scaling across them [20]. However, due to narrow overall dynamic power range of servers in a data center, it has been found beneficial to consolidate workload to a limited number of servers and switch off or put to sleep / hibernate state idle nodes [21].

Another technology that can improve the utilization of resources, and thus reduce the power consumption is virtualization of computer resources. Virtualization technology allows one to create several Virtual Machines (VMs) on a physical server and, therefore, reduce the amount of hardware in use and improve the utilization of resources. The concept originated with the IBM mainframe operating systems of the 1960s, but was commercialized for x86-compatible computers only in the 1990s. Several commercial companies and open-source projects now offer software packages to enable a transition to virtual computing. Intel Corporation and AMD have also built proprietary virtualization enhancements to the x86 instruction set into each of their CPU product lines, in order to facilitate virtualized computing. Among the benefits of virtualization are improved fault and performance isolation between applications sharing the same computer node (a VM is viewed as a dedicated resource to the customer); the ability to relatively easily move VMs from one physical host to another using live or off-line migration; and support for hardware and software heterogeneity. The ability to reallocate VMs in run-time enables dynamic consolidation of the workload, as VMs can be moved to a minimal number of physical nodes, while idle nodes can be switched to power saving modes.

Terminal servers have also been used in Green IT practices. When using terminal servers, users connect to a central server; all of the computing is done at the server level but the end user experiences a dedicated computing resource. It is usually combined with thin clients, which use up to 1/8 the amount of energy of a normal workstation, resulting in a decrease of the energy consumption and costs. There has been an increase in the usage of terminal services with thin clients to create virtual laboratories. Examples of terminal server software include Terminal Services for Windows, the Aqua Connect Terminal Server for Mac, and the Linux Terminal Server Project (LTSP) for the Linux operating system. Thin clients possibly are going to gain a new wave of popularity with the adoption of the Software as a Service (SaaS) model, which is one of the kinds of Cloud computing [22], or Virtual Desktop Infrastructures (VDI) heavily promoted by virtualization software vendors[5].

Traditionally, an organization purchases its own computing resources and deals with the maintenance and upgrades of the outdated hardware and software, resulting in additional expenses. The recently emerged Cloud computing paradigm [22] leverages virtualization technology and provides the ability to provision resources on-demand on a pay-as-you-go basis. Organizations can outsource their computation needs to the Cloud, thereby eliminating the necessity to maintain own computing infrastructure. Cloud computing naturally leads to power-efficiency by providing the following characteristics:

---

[3] Advanced power management. http://en.wikipedia.org/wiki/Advanced_power_management
[4] Advanced Configuration & Power Interface. http://www.acpi.info/
[5] VMware View (VMware VDI) Enterprise Virtual Desktop Management. http://www.vmware.com/products/view/
  Citrix XenDesktop Desktop Virtualization. http://www.citrix.com/virtualization/desktop/xendesktop.html
  Sun Virtual Desktop Infrastructure Software. http://www.sun.com/software/vdi/

- Economy of scale due to elimination of redundancies.
- Improved utilization of the resources.
- Location independence – VMs can be moved to a place where energy is cheaper.
- Scaling up and down – resource usage can be adjusted to current requirements.
- Efficient resource management by the Cloud provider.

One of the important requirements for a Cloud computing environment is providing reliable QoS. It can be defined in terms of Service Level Agreements (SLA) that describe such characteristics as minimal throughput, maximal response time or latency delivered by the deployed system. Although modern virtualization technologies can ensure performance isolation between VMs sharing the same physical computing node, due to aggressive consolidation and variability of the workload some VMs may not get the required amount of resource when requested. This leads to performance losses in terms of increased response time, time outs or failures in the worst case. Therefore, Cloud providers have to deal with the power-performance trade-off – minimization of the power consumption, while meeting the QoS requirements.

The following sections detail different levels of the presented taxonomy: in Section 5 we will discuss power optimization techniques that can be applied at the hardware level. We will consider the approaches proposed for power management at the operating system level in Section 6, followed by the discussion of modern virtualization technologies and their impact on power-aware resource management in Section 7 and the recent approaches applied at the data center level in Section 8.

# 5 Hardware and Firmware Level

As shown in Figure 5, DPM techniques applied at the hardware and firmware level can be broadly divided into two categories: Dynamic Component Deactivation (DCD) and Dynamic Performance Scaling (DPS). DCD techniques are built upon the idea of the clock gating of parts of an electronic component or complete disabling during periods of inactivity.

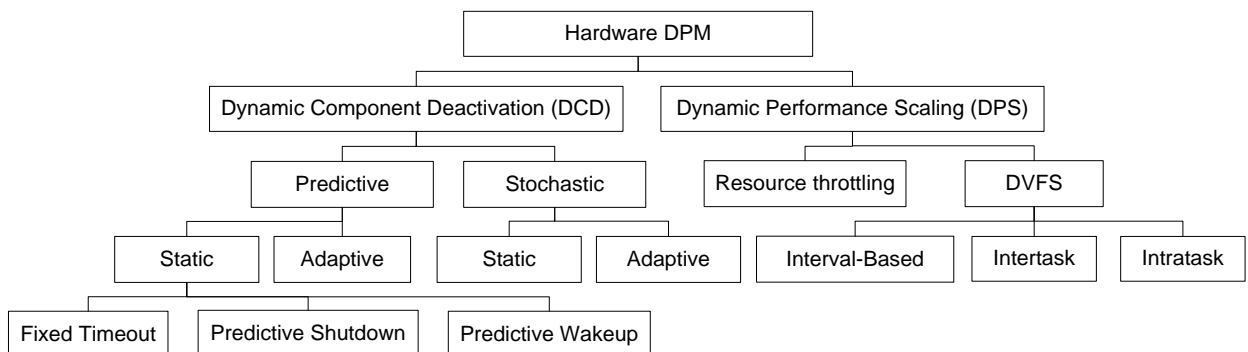

Figure 5. DPM techniques applied at the hardware and firmware levels.

The problem could be easily solved if transitions between power states would cause negligible power and performance overhead. However, transitions to low-power states usually lead to additional power consumption and delays caused by the re-initialization of the components. For example, if entering a low-power state requires shut-down of the power supply, returning to the active state will cause a delay consisting of: turning on and stabilizing the power supply and clock; re-initialization of the system; and restoring the context [23]. In the case of non-negligible transitions, efficient power management turns into a difficult on-line optimization problem. A transition to low-power state is worthwhile only if the period of inactivity is longer than the aggregated delay of transitions from and into the active state, and saved power is higher than required to reinitialize the component.

## 5.1 Dynamic Component Deactivation (DCD)

Computer components that do not support performance scaling and can only be deactivated require techniques that will leverage the workload variability and disable the component when it is idle. The problem is trivial in the case of a negligible transition overhead. However, in reality such transitions lead not only to delays, which can degrade performance of the system, but to additional power draw. Therefore, to achieve efficiency a transition has to be done only if the idle period is long enough to cover the transition overhead. In most real-world systems there is a limited or no knowledge about the future workload. Therefore, a prediction of an effective transition has to be done according to historical data or some system model. A large volume of research has been done to develop efficient methods to solve this problem [23] [24]. As shown in Figure 5, the proposed DCD techniques can be divided into predictive and stochastic.

Predictive techniques are based on the correlation between the past history of the system behavior and its near future. The efficiency of such techniques is highly dependent on the actual correlation between past and future events and quality of tuning for a particular type of the workload. A non-ideal prediction can result in an over-prediction or under-prediction. An over-prediction means that the actual idle period is shorter than the predicted leading to a performance penalty. On the other hand, an under-prediction means that the actual idle period is longer the predicted. This case does not have any influence on the performance; however, it results in reduced energy savings. Predictive techniques can be further split into static and adaptive, which are discussed below.

Static techniques utilize some threshold for a real-time execution parameter to make predictions of idle periods. The simplest policy is called fixed timeout. The idea is to define the length of time after which a period of inactivity can be treated as long enough to do a transition to a low-power state. Activation of the component is initiated once the first request to a component is received. The policy has two advantages: it can be applied to any workload type, and over- and under-predictions can be controlled by adjusting the value of the timeout threshold. However, disadvantages are obvious: the policy requires adjustment of the threshold value for each workload, it always leads to a performance loss on the activation, and the energy consumed from the beginning of an idle period to the timeout is wasted. Two ways to overcome the drawbacks of the fixed timeout policy have been proposed: predictive shutdown and predictive wakeup. Predictive shutdown policies address the problem of the missed opportunity to save energy within the timeout. These policies utilize the assumption that previous periods of inactivity are highly correlated with the nearest future. According to the analysis of the historical information they predict the length of the next idle period before it actually starts. These policies are highly dependent on the actual workload and strength of the correlation between past and future events. History-based predictors have been shown to be more efficient and less safe than timeouts [25]. Predictive wakeup techniques aim to eliminate the performance penalty on the activation. The transition to the active state is predicted based on the past history and performed before an actual user request [26]. This technique increases the energy consumption, but reduces performance losses on wakeups.

All the static techniques are inefficient in cases when the system workload is unknown or can vary over time. To address this problem adaptive predictive techniques have been introduced. The basic idea is to dynamically adjust the parameters, which are fixed for the static techniques, according to the quality of prediction that they have provided in the past. For example, the timeout value can be increased if for the last several intervals the value has lead to over-prediction. Another way to provide the adaptation is to maintain a list of possible values of the parameter of interest and assign weights to the values according to their efficiency at previous intervals. The actual value is obtained as a weighted average over all the values in the list. In general, adaptive techniques are more efficient than static when the type of the workload is unknown a priori. Several adaptive techniques are discussed in the paper by Douglis et al. [27].

Another way to deal with non-deterministic system behavior is to formulate the problem as a stochastic optimization, which requires building of an appropriate probabilistic model of the system. For instance, in such a model system requests and power state transitions are represented as stochastic processes and can be modelled as Markov processes. At any moment, a request arrives with some probability and a device power state transition occurs with another probability obtained by solving the stochastic optimization problem. It is important to note, that the results, obtained using the stochastic approach, are expected values, and there is no guarantee that the solution will be optimal for a particular case. Moreover, constructing a stochastic model of the system in practice may not be straightforward. If the model is not accurate, the policies using this model may not provide an efficient system control.

## 5.2 Dynamic Performance Scaling (DPS)

Dynamic Performance Scaling (DPS) includes different techniques that can be applied to computer components supporting dynamic adjustment of their performance proportionally to the power consumption. Instead of complete deactivations, some components, such as CPU, allow gradual reductions or increases of the clock frequency along with the adjustment of the supply voltage in cases when the resource is not utilized for the full capacity. This idea lies in the roots of the widely adopted Dynamic Voltage and Frequency Scaling (DVFS) technique.

### 5.2.1 Dynamic Voltage and Frequency Scaling (DVFS)

Although the CPU frequency can be adjusted separately, frequency scaling by itself is rarely worthwhile as a way to conserve switching power. Saving the most power requires dynamic voltage scaling too, because of the $V^2$ component and the fact that modern CPUs are strongly optimized for low voltage states. Dynamic voltage scaling is usually used in conjunction with frequency scaling, as the frequency that a chip may run at is related to the operating voltage. The efficiency of some electrical components, such as voltage regulators, decreases with a temperature increase, so the power used may increase with temperature. Since increasing power use may raise the temperature, increases in voltage or frequency may raise the system power demand even faster than the CMOS formula indicates, and vice-versa. DVFS reduces the number of instructions a processor can issue in a given amount of time, thus reducing the performance. This, in turn, increases run time for program segments which are sufficiently CPU-bound. Hence, it creates challenges of providing optimal energy / performance control, which have been extensively investigated by scientists in recent years. Some of the research works will be reviewed in the following sections.

Although the application of DVFS may seem to be straightforward, real-world systems raise many complexities that have to be considered. First of all, due to complex architectures of modern CPUs (i.e. pipelining, multi-level cache, etc.), the prediction of the required CPU clock frequency that will meet application's performance requirements is not trivial. Another problem is that in contrast to the theory, power consumption by a CPU may not be quadratic to its supply voltage. For example, in [8] it is shown that some architectures may include several supply voltages that power different parts of the chip, and even if one of them can be reduced, overall power consumption will be dominated by the larger supply voltage. Moreover, execution time of the program running on the CPU may not be inversely proportional to the clock frequency, and DVFS may result in non-linearities in the execution time [28]. For example, if the program is memory or I/O bounded, CPU speed will not have a dramatic effect on the execution time. Furthermore, slowing down the CPU may lead to changes in the order in which tasks are scheduled [8]. In summary, DVFS can provide substantial energy savings; however, it has to be applied carefully, as the result may significantly vary for different hardware and software system architectures.

Approaches that apply DVFS to reduce energy consumption by a system can be divided into interval-based, intertask and intratask [28]. Interval-based algorithms are similar to adaptive predictive DCD approaches in that they also utilize knowledge of the past periods of the CPU activity [29] [30]. Depending on the utilization of the CPU during previous intervals, they predict the utilization in the near future and appropriately adjust the voltage and clock frequency. Wierman et al. [31] and Andrew et al. [32] have conducted analytical studies of speed scaling algorithms in processor sharing systems. They have proved that no online energy-proportional speed scaling algorithm can be better than 2-competitive comparing to the offline optimal algorithm. Moreover, they have found that sophistication in the design of speed scaling algorithms does not provide significant performance improvements; however, it dramatically improves robustness to errors in estimation of workload parameters. Intertask approaches instead of relying on coarse grained information on the CPU utilization, distinguish different tasks running in the system and assign them different speeds [33] [34]. The problem is easy to solve if the workload is known a priori or constant over all the period of a task execution. However, the problem becomes non-trivial when the workload is irregular. In contrast to intertask, intratask approaches leverage fine grained information about the structure of programs and adjust the processor frequency and voltage within the tasks [35] [36]. Such policies can be implemented by splitting a program execution into timeslots and assigning different CPU speeds to each of them. Another way is to implement them at the compiler level. This kind of approaches utilizes compiler's knowledge of a program's structure to make inferences about possible periods for the clock frequency reduction.

## 5.3 Advanced Configuration and Power Interface

Many DPM algorithms, such as timeout-based as well as other predictive and stochastic policies, can be implemented in the hardware as a part of an electronic circuit. However, a hardware implementation highly complicates the modification and reconfiguration of the policies. Therefore, there are strong reasons to shift the implementation to the software level. In 1996 to address this problem Intel, Microsoft and Toshiba have published the first version of the Advanced Configuration and Power Interface (ACPI) specification – an open standard defining a unified operating system-centric device configuration and power management interface. In contrast to previous BIOS central, firmware-based and platform specific power management systems, ACPI describes platform-independent interfaces for hardware discovery, configuration, power management and monitoring.

ACPI is an attempt to unify and improve the existing power and configuration standards for hardware devices. The standard brings DPM into the operating system control and requires an ACPI-compatible operating system to take over the system and have the exclusive control of all aspects of the power management and device configuration responsibilities. The main goals of ACPI are to enable all computing systems to implement dynamic power management capabilities, and simplify and accelerate the development of power-managed systems. It is important to note that ACPI does not put any constraints on particular power management policies, but provides the interface that can be used by software developers to leverage flexibility in adjustment of the system's power states.

ACPI defines a number of power states that can be applied in the system in run-time. The most important states in the context of DPM are C-states and P-states. C-states are the CPU power states C0-C3 that denote the operating state, halt, stop-clock and sleep mode accordingly. While the processor operates, it can be in one of several power-performance states (P-state). Each of these states designates a particular combination of DVFS settings. P-states are implementation-dependent, but P0 is always the highest-performance state, with $P_1$ to $P_n$ being successively lower-performance states, up to an implementation-specific limit of *n* no greater than 16. P-states have become known as SpeedStep in Intel processors, PowerNow! or Cool'n'Quiet in AMD processors, and PowerSaver in VIA processors. ACPI is widely used by operating systems, middleware and software on top of them to manage power consumption according to their specific policies.

# 6 Operating System Level

In this section, we will discuss research works that deal with power efficient resource management at the operating system level. The taxonomy of the characteristics used to classify the works is presented in Figure 6. To highlight the most important characteristics of the works, they are summarized in Table 2 (full table is given in Appendix A).

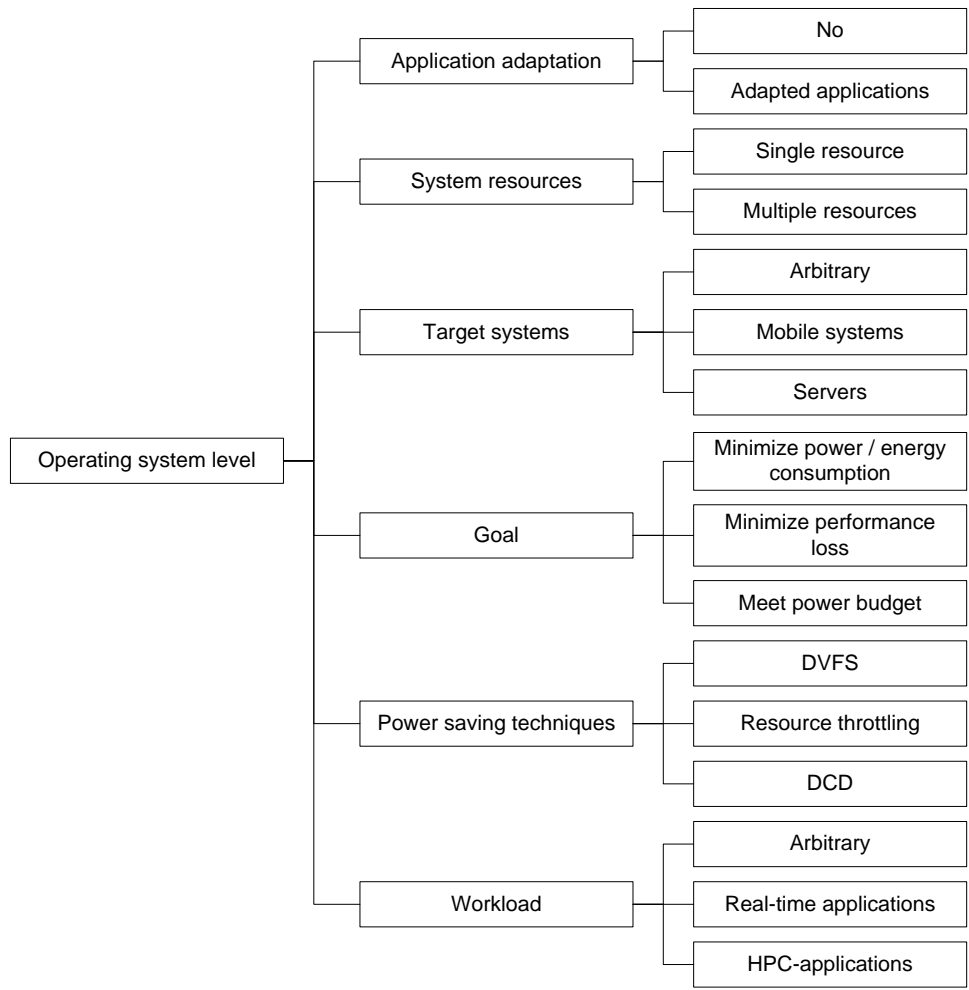

Figure 6. Operating system level taxonomy

Table 2. Operating system level research works.

| Project name | System resources | Target systems | Goal | Power-saving techniques |
|---|---|---|---|---|
| The Ondemand Governor, Pallipadi and Starikovskiy [19] | CPU | Arbitrary | Minimize power consumption, minimize performance loss | DVFS |
| ECOsystem, Zeng et al. [37] [38] | CPU, memory, disk storage, network interface | Mobile systems | Achieve target battery lifetime | Resource throttling |

| Project name | System resources | Target systems | Goal | Power-saving techniques |
|---|---|---|---|---|
| Nemesis OS, Neugebauer and McAuley [39] | CPU, memory, disk storage, network interface | Mobile systems | Achieve target battery lifetime | Resource throttling |
| GRACE, Sachs et al. [40] [41] | CPU, network interface | Mobile systems | Minimize energy consumption, satisfy performance requirements | DVFS, resource throttling |
| Linux/RK, Rajkumar et al. [42] | CPU | Real-time systems | Minimize energy consumption, satisfy performance requirements | DVFS |
| Coda and Odyssey, Flinn and Satyanarayanan [43] | CPU, network interface | Mobile systems | Minimize energy consumption by application degradation | Resource throttling |
| PowerNap, Meisner et al. [44] | System-wide | Server systems | Minimize power consumption, minimize performance loss | DCD |

## 6.1 The Ondemand Governor (Linux Kernel)

Pallipadi and Starikovskiy [19] have developed an in-kernel real-time power manager for Linux OS called the ondemand governor. The manager continuously monitors the CPU utilization multiple times per second and sets a clock frequency and supply voltage pair that corresponds to current performance requirements keeping the CPU approximately 80% busy to handle fast changes in the workload. The goal of the ondemand governor is to keep the performance loss due to reduced frequency to the minimum. Modern CPU frequency scaling technologies provides extremely low latency allowing dynamic adjustment of the power consumption matching the variable workload with almost negligible performance overhead. For example, Enhanced Intel Speedstep Technology enables frequency switching with the latency as low as 10 ms. To accommodate to different requirements of diverse systems, the ondemand governor can be tuned via specification of the rate at which the CPU utilization is checked and upper utilization threshold, which is set to 80% by default.

The ondemand governor effectively handles multiprocessor SMP systems, as well as multi-core and multi-threading CPU architectures. The governor manages each CPU individually and can manage different cores in the CPU separately if this is supported by the hardware. In cases if different processor cores in a CPU are dependent on each other in terms of frequency, they are managed together as a single entity. In order to support this design, the ondemand governor will set the frequency to all of the cores based on the highest utilization among the cores in the group.

There are a number of improvements that are currently under investigation, including parallel calculation of the utilization and a dedicated work queue. The original governor samples the utilization of all of the processors in the system in a centralized way that can become a significant overhead with increase of the number of CPUs. To overcome this problem the authors have proposed a parallel sampling independently for each CPU. Another improvement that can increase performance for multiprocessor systems is to have dedicated kernel threads for the governor and do sampling and changing of frequencies in the context of a particular kernel thread.

## 6.2 ECOsystem

Zeng et al. [37] [38] have proposed and developed ECOsystem – a framework for managing energy as a first-class OS resource aimed at battery powered devices. The authors' fundamental assumption is that applications play an important role in energy distribution opportunities that can be leveraged only at the application level. ECOsystem provides an interface to define a target battery lifetime and applications' priorities used to determine the amount of energy that will be allocated to applications at each time frame.

The authors split OS-level energy management into two dimensions. Along the first dimension, there are a variety of the system devices (e.g. CPU, memory, disk storage, network interface) that can consume energy concurrently. The other dimension spans applications that share the system devices and cause the energy consumption. To address the problem of accounting the energy usage by both devices and applications, the authors have introduced a new measurement unit called currentcy. One unit of currentcy represents the right to consume a certain amount of energy during a fixed period of time. When the user sets the target battery lifetime and prioritises the applications, ECOsystem transforms these data into an appropriate amount of currentcy and determines how much currentcy should be allocated to each application at each time frame. The length of the timeframe has been empirically determined as 1 second that is sufficient to achieve smooth energy allocation. An application expends the allocated amount of currentcy by utilizing the CPU, performing disk and memory accesses and consuming other system resources. An application can accumulate currentcy up to a specified limit. When an expenditure of an application exceeds the allocated amount of currentcy, none of the associated processes are scheduled or otherwise serviced.

The system has been implemented as a modified Linux kernel and has been experimentally evaluated. The obtained results show that the proposed model can be effectively used to meet different energy goals, such as achieving a target battery lifetime and proportional energy distribution among competing applications.

## 6.3 Nemesis OS

Neugebauer and McAuley [39] have developed the resource-centric Nemesis OS – an operating system for battery powered devices that strive to provide a consistent QoS for time-sensitive application, such as multimedia applications. Nemesis provides fine grained control and accounting for energy usage over all system resources: CPU, memory, disk and network bandwidth.

To implement per-process resource usage accounting, the OS has been vertically structured: most of the system's functions, protocol stacks and device drivers are implemented in user-level shared libraries that execute in the applications' processes. This design allows accurate and easy accounting for the energy consumption caused by individual applications.

The goal of Nemesis is to address the problem of battery lifetime management. To achieve the target battery lifetime specified by the user, the system relies on the cooperation with applications. If the current energy consumption rate exceeds the threshold that can lead to failing to meet the user's expectations, the system charges the applications according to their current energy usage. The applications should interpret the charges as feedback signals and adapt their behavior. The applications are supposed to limit their resource usage according to the data provided by the OS. However, not all application may support the adaptation. In this case the user can prioritise the applications leading to shut down of the low-priority tasks. Nemesis currently supports a number of platforms including Intel 486, Pentium, Pentium Pro and Pentium II based PCs, DEC Alpha workstations and evaluation boards, and StrongARM SA-110 based network computers.

## 6.4 The Illinois GRACE project

Sachs et al. [40] [41] have developed the Illinois GRACE project (Global Resource Adaptation through CoopEration). They have proposed saving energy through coordinated adaptation at multiple system layers according to changes in the applications' demand for system resources. The authors have proposed three levels of adaptation: global, per-application and internal adaptation. The global adaptation takes into account all the applications running in the system and all the system layers. This level of adaptation responses to significant changes in the system, such as application entry or exit. The per-application adaptation considers each application in isolation and is invoked every time frame adapting all the system resources to the application's demands. The internal adaptation focuses on different system resources separately that are possibly shared by multiple applications and adapts the states of the resources. All the adaptation levels are coordinated in order to ensure adaptation decisions that are effective across all levels.

The framework supports adaptations of the CPU performance (DVSF), applications (frame rate and dithering), and soft CPU scaling (CPU time allocation). The second generation of the framework (GRACE-2) focuses on a hierarchical adaptation for mobile multimedia systems. Moreover, it leverages the adaptation of the application behavior depending on the resource constraints. GRACE-2 apart from the CPU adaptation enforces network bandwidth constraints and minimizes network transmission energy. The approach has been implemented as a part of the Linux kernel and requires applications to be able to limit their resource usage in run-time on order to leverage the per-application adaptation technique. There is only a limited support for legacy applications.

The experimental results show that the application adaptation provides significant benefits over the global adaptation when the network bandwidth is constrained. Energy savings in a system with the CPU and network adaptations when adding the application adaptation reach 32% (22% on average). When both the CPU and application adaptations are added to a system with the global adaptation, the energy savings have been found to be more than additive.

## 6.5 Linux/RK

Rajkumar et al. [42] have proposed several algorithms for application of DVFS in real-time systems and have implemented a prototype as a modified Linux kernel – Linux/Resource Kernel (Linux/RK). The objective is to minimize the energy consumption, while maintaining the performance isolation between applications. The authors have proposed four alternative DVFS algorithms that are automatically selected by the system when appropriate.

SystemClock Frequency Assignment (Sys-Clock) is suitable for systems where the overhead of voltage and frequency scaling is too high to perform at every context switch. A single clock frequency is selected at the admission of an application and kept constant until a set of applications running in the system changes. Priority-Monotonic Clock Frequency Assignment (PM-Clock) is suitable for systems with a low voltage and frequency scaling overhead allowing adjustment of the voltage and frequency settings at each context switch. Each application is assigned its own constant clock frequency which is enabled when the application is allocated a CPU time frame. Optimal Clock Frequency Assignment (Opt-Clock) uses a non-linear optimisation model to determine an optimal frequency for each application that minimizes the energy consumption. Due to high computational complexity this technique is suitable only for offline usage. Dynamic PM-Clock (DPM-Clock) suits systems where the average execution time of an application is significantly less than the worst case. The authors have conducted experimental studies to evaluate the proposed algorithms. The results show that SysClock, PM-Clock and DPM-Clock provide up to 50% energy savings.

## 6.6 Coda and Odyssey

Flinn and Satyanarayanan [43] have explored the problem of managing limited computing resources and battery lifetime in mobile systems, as well as addressing the variability of the network connectivity. They have developed two systems, Coda and Odyssey that implement adaptation across multiple system levels. Coda implements application-transparent adaptation in the context of a distributed file system, which does not require any modification of legacy applications to run in the system.

Odyssey is responsible for initiation and managing of application adaptations. This kind of adaptation allows adjustment of the resource consumption by the cost of the output data quality, which is mostly suitable for multimedia applications. For example, video data can be processed or transferred over network in lower resolution or sound quality can be reduced in cases of constrained resources.

Odyssey introduces a term fidelity that defines the degree to which the output data corresponds to the original quality. Each application can specify acceptable levels of fidelity that can be requested by Odyssey when the resource usage has to be limited. When Odyssey notifies an application about a change of the resource availability, the application has to adjust its fidelity to match the requested level. For energy-aware adaptation it is essential that reductions in fidelity lead to energy savings that are both significant and predictable. The evaluation results show that this approach allows the extension of the battery lifetime up to 30%. A limitation of such a system is that all the necessary applications have to be modified in order to support the proposed approach.

## 6.7 PowerNap

Meisner et al. [44] have proposed an approach for power conservation in server systems based on fast transitions between active and low power states. The goal is to minimize power consumption by a server while it is in an idle state. Instead of addressing the problem of achieving energy-proportional computing as proposed by Barroso and Holzle [9], the authors require only two power states (sleep and fully active) for each system component. The other requirements are fast transitions between the power states and very low power consumption in the sleep mode.

To investigate the problem, the authors have collected fine grained utilization traces of several servers serving different workloads. According to the data, the majority of idle periods are shorter than 1 second with the mean length in the order of hundreds of milliseconds. Whereas, busy periods are even shorter falling bellow 100 ms for some workloads. The main idea of the proposed approach is to leverage short idle periods that occur due to the workload variability. To estimate the characteristics of the hardware able to implement the technique, the authors have constructed a queueing model based on characteristics of the collected utilization traces. They have found that if the transition time is less than 1 ms, it becomes negligible and power savings vary linearly with the utilization for all workloads. However, with the growth of the transition time, power savings decrease and the performance penalty becomes higher. When the transition time reaches 100 ms, the relative response time for low utilization can grow up to 3.5 times relatively to a system without power management, which is clearly unacceptable for real-world systems.

The authors have concluded that if the transition time is less than 10 ms, power savings are approximately linear to the utilization and significantly outperform the effect from DVFS for low utilization (less than 40%). However, the problem is that the requirement for the transition time being less than 10 ms cannot be satisfied by the current level of technology. According to the data provided by the authors, modern servers can ensure the transition time of 300 ms, which is anyway far from the requested 10 ms. The proposed approach is similar to the fixed time-out DCD technique, but adapted to fine grained management. Therefore, all the disadvantages of the fixed time-out technique are inherited by the proposed approach, i.e. constant performance penalty on wake ups and the overhead in cases when an idle period is shorter then the transition time to and from a low power state. The authors have reported that if the stated requirements are satisfied, the average server power consumption can be reduced by 74%.

# 7 Virtualization Level

The virtualization level enables the abstraction of an OS and applications running on it from the hardware. Physical resources can be split into a number of logical slices called Virtual Machines (VMs). Each VM can accommodate an individual OS creating for the user a view of a dedicated physical resource and ensuring performance and failure isolation between VMs sharing a single physical machine. The virtualization layer lies between the hardware and OS and; therefore, a Virtual Machine Monitor (VMM) takes control over resource multiplexing and has to be involved in the system's power management in order to provide efficient operation. There are two ways of how a VMM can participate in the power management:

1. A VMM can act as a power-aware OS without distinction between VMs: monitor the overall system's performance and appropriately apply DVFS or any DCD techniques to the system components.
2. Another way is to leverage OS's specific power management policies and application-level knowledge, and map power management calls from different VMs on actual changes in the hardware's power state or enforce system-wide power limits in a coordinated manner.

We will discuss these techniques in detail in the following sections.

## 7.1 Virtualization Technology Vendors

In Section 7.1 we will discuss three of the most popular virtualization technology solutions: the Xen hypervisor[6], VMware solutions[7] and KVM[8]. Both of these systems support the first described way to perform power management, however, neither allows coordination of VMs' specific calls for power state changes. Section 7.2 discusses an approach proposed by Stoess et al. [45] that utilizes both system-wide power control and fine grained application-specific power management performed by guest operating systems.

Other important capabilities supported by the mentioned virtualization solutions are offline and live migrations of VMs. They enable transferring VMs from one physical host to another, and thus have facilitated the development of different techniques for virtual machines consolidation and load balancing that will be discussed in Section 8.

### 7.1.1 Xen

The Xen hypervisor is an open source virtualization technology developed collaboratively by the Xen community and engineers from over 20 innovative data center solution vendors [46]. Xen is licensed under the GNU General Public License (GPL2) and available at no charge in both source and object formats. Xen's support for power management is similar to what is provided by the Linux's ondemand governor described in Section 6.1. Xen supports ACPI's P-states implemented in the cpufreq driver [47]. The system periodically measures the CPU utilization, determines an appropriate P-state and issue a platform-dependent command to make a change in the hardware's power state. Similarly to the Linux's power management subsystem, Xen provides four governors:

---

[6] http://www.xen.org/
[7] http://www.vmware.com/
[8] http://www.linux-kvm.org/

- Ondemand – chooses the best P-state according to current resource requirements.
- Userspace – sets the CPU frequency specified by the user.
- Performance – sets the highest available clock frequency.
- Powersave – sets the lowest clock frequency.

Apart from P-states, Xen also incorporates the support for C-states (CPU sleeping states) [47]. When a physical CPU does not have any task assigned, it is switched to a C-state. When a new request comes, the CPU is switched back to the active state. An issue is to determine which C-state to enter: deeper C-states provide higher energy saving by the cost of higher transition latency. At this moment, by default Xen puts the CPU into the first C-state, which provides the least transition delay. However, the user can specify a C-state to enter. As the CPU wakes up upon receiving a load, it always gets an inevitable performance penalty. The policy is a fixed timeout DCD implying all its disadvantages described in Section 5.1.

Besides P- and C-states, Xen also supports regular and live migration of VMs, which can be leveraged by power-aware dynamic VM consolidation algorithms. Migration is used to transfer a VM between physical hosts. Regular migration moves a VM from one host to another by suspending, copying the VM's memory contents, and then resuming the VM on the destination host. Live migration allows transferring a VM without suspension and from the user side the migration should be inconspicuous. To perform a live migration, both hosts must be running Xen and the destination host must have sufficient resources (e.g. memory capacity) to accommodate the VM after the transmission. Xen starts a new VM instance that forms a container for the VM to be migrated. Xen cyclically copies memory pages to the destination host, continuously refreshing the pages that have been updated on the source. When it notices that the number of modified pages is not shrinking anymore, it stops the source instance and copies the remaining memory pages. Once it is completed, the new VM instance is started. To minimize the migration overhead, the hosts are usually connected to a Network Attached Storage (NAS) or similar storage solution, which eliminates the necessity to copy disk contents. The developers argue that the final phase of a live migration, when both instances are suspended, typically takes approximately 50 ms. Given such a low overhead, the live migration technology has facilitated the development of various energy conservation dynamic VM consolidation approaches proposed by researchers around the world.

### 7.1.2 VMware

VMware ESX Server and VMWare ESXi are enterprise-level virtualization solutions offered by VMware, Inc. Similarly to Xen, VMware supports host-level power management via DVFS. The system monitors the CPU utilization and continuously applies appropriate ACPI's P-states [48]. VMware VMotion and VMware Distributed Resource Scheduler (DRS) are two other services that operate in conjunction with ESX Server and ESXi [49]. VMware VMotion enables live migration if VMs between physical nodes. A migration can be initiated manually by system administrators or programmatically. VMware DRS monitors the resource usage in a pool of servers and uses VMotion to continuously rebalance VMs according to the current workload and load balancing policy.

VMware DRS contains a subsystem called VMware Distributed Power Management (DPM) to reduce power consumption by a pool of servers by dynamically switching off spare servers [49] [50]. Servers are powered back when there is a rising demand for the resource capacity. VMware DPM utilizes live migration to reallocate VMs keeping the minimal number of servers powered on. VMware ESX Server and VMware ESXi are free for use, whereas other components of VMware Infrastructure have a commercial license.

### 7.1.3 KVM

KVM is a virtualization platform, which is implemented as a module of the Linux kernel [51]. Under this model Linux works as a hypervisor, and all the VMs are regular processes scheduled by the Linux scheduler. This approach reduces the complexity of the hypervisor implementation, as scheduling and memory management are handled by the Linux kernel.

KVM supports the S4 (hibernate) and S3 (sleep / stand by) power states[9]. S4 does not require any specific support from KVM: on hibernation the guest OS dumps memory state to a hard disk and initiates powering off the computer. The hypervisor translates this signal into termination of the appropriate process. On the next boot, the OS reads the saved memory state from the disk, resumes from the hibernation and reinitializes all the devices. During the S3 state memory is kept powered, and thus the content does need to be saved to a disk. However, the guest OS must save states of the devices, as they should be restored on resume. During the next boot, the BIOS should recognize the S3 state and not initialize the devices, but jump directly to the restoration of the saved device states. Therefore, the BIOS is modified in order to support such behaviour.

## 7.2 Energy Management for Hypervisor-Based Virtual Machines

Stoess et al. [45] have proposed a framework for energy management in virtualized servers. Typically, energy-aware OSes assume full knowledge and full control over the underlying hardware, implying device- or application level accounting for the energy usage. However, in virtualized systems, a hardware resource is shared among multiple VMs. In such an environment, device control and accounting information are distributed across the whole system, making it infeasible for an OS to take a full control over the hardware. This results in inability of energy-aware OSes to invoke their policies in the system. The authors have proposed mechanisms for fine grained guest OS-level energy accounting and allocation. To encompass the diverse demands on energy management, the authors have proposes to use the notion of energy as the base abstraction in the system, an approach similar to the currentcy model in ECOsystem described in Section 6.2.

The prototypical implementation comprises two sub-systems: a host-level resource manager and an energy-aware OS. The host-level manager enforces system-wide power limits across VM instances. The power limits can be dictated by a battery or power generator, or by thermal constraints imposed by reliability requirements and the cooling system capacity. The manager determines power limits for each VM and device type, which cannot be exceeded to meet the defined power constraints. The complementary energy-aware OS is capable of fine grained application-specific energy management. To enable application-specific energy management, the framework supports accounting and control not only for physical but also of virtual devices. This enables guest resource management subsystems to leverage their application-specific knowledge.

Experimental results presented by the authors show that the prototype is capable of enforcing power limits for energy-aware and energy-unaware guest OSes. Three areas are considered to be prevalent for future work: devices with multiple power states, processors with support for hardware-assisted virtualization, and multi-core architectures.

# 8 Data Center Level

In this chapter recently proposed approaches that deal with power management at the data center level are discussed. The characteristics used to classify the approaches are presented in Figure 7.

---

[9] http://www.linux-kvm.org/page/PowerManagement

Usually an approach is based on consolidation of the workload across physical nodes in data centers. The workload can be represented by incoming requests for online services or web applications, or virtual machines. The goal is to allocate requests / virtual machines to the minimal amount of physical resources and turn off or put in sleep / hibernate state the idle resources. The problem of the allocation is twofold: firstly, it is necessary to allocate new requests; secondly, the performance of existing applications / VMs should be continuously monitored and if required the allocation should be adapted to achieve the best possible power-performance trade-off regarding to specified QoS.

Table 3 illustrates the most significant characteristics of the reviewed research works (full table is given in Appendix B).

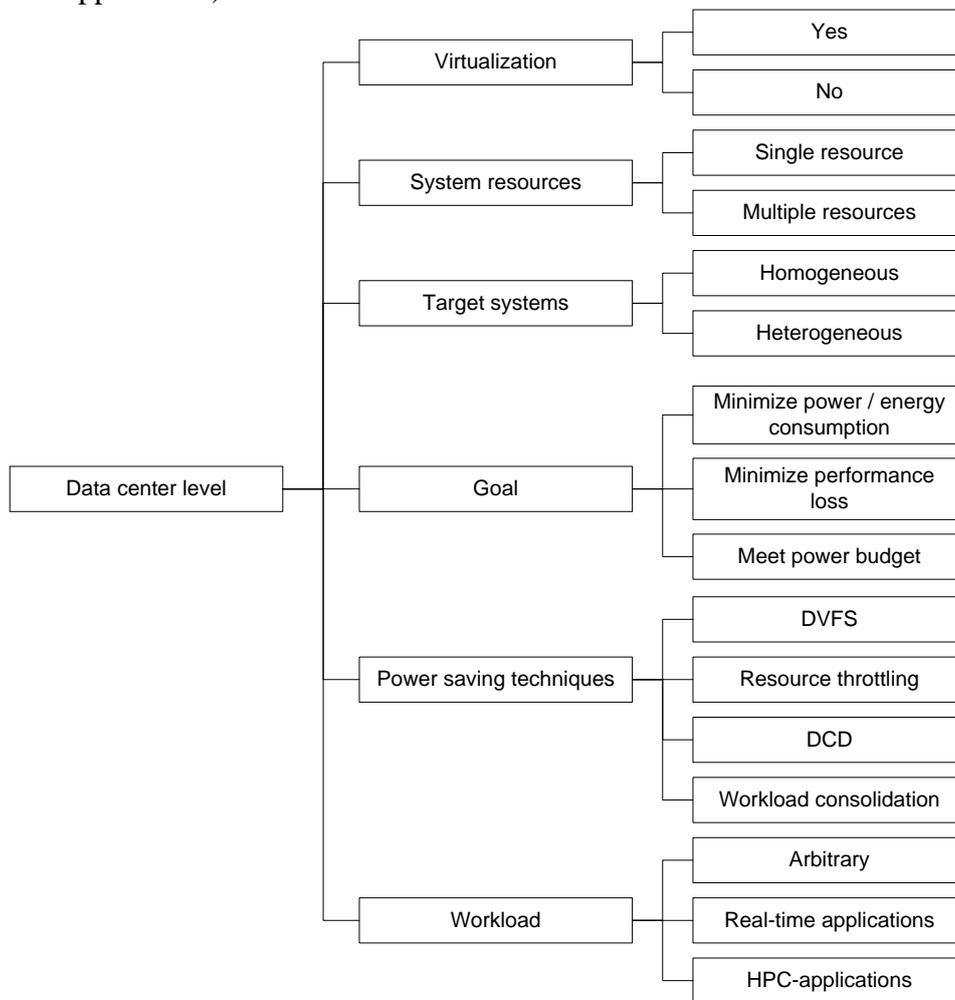

Figure 7. Data center level taxonomy

Table 3. Data center level research works.

| Project name | Virtua-lization | System resources | Goal | Power-saving techniques |
|---|---|---|---|---|
| Load Balancing and Unbalancing for Power and Performance in Cluster-Based System, Pinheiro et al. [21] | No | CPU, disk storage, network interface | Minimize power consumption, minimize performance loss | Server power switching |
| Managing Energy and Server Resources in Hosting Centers, Chase et | No | CPU | Minimize power consumption, minimize performance loss | Workload consolidation, server power |

| Project name | Virtua-lization | System resources | Goal | Power-saving techniques |
|---|---|---|---|---|
| al. [52] | | | | switching |
| Energy-Efficient Server Clusters, Elnozahy et al. [20] | No | CPU | Minimize energy consumption, satisfy performance requirements | DVFS, server power switching |
| Energy-Aware Consolidation for Cloud Computing, Srikantaiah et al. [53] | No | CPU, disk storage | Minimize energy consumption, satisfy performance requirements | Workload consolidation, server power switching |
| Optimal Power Allocation in Server Farms, Gandhi et al. [54] | No | CPU | Allocate the available power budget to minimize mean response time | DVFS |
| Environment-Conscious Scheduling of HPC Applications, Garg et al. [55] | No | CPU | Minimize energy consumption and $CO_2$ emissions, maximize profit | DVFS, leveraging geographical distribution of data centers |
| VirtualPower: Coordinated Power Management in Virtualized Enterprise Systems, Nathuji and Schwan [56] | Yes | CPU | Minimize energy consumption, satisfy performance requirements | DFVS, soft scaling, VM consolidation, server power switching |
| Coordinated Multi-level Power Management for the Data Center, Raghavendra et al. [57] | Yes | CPU | Minimize power consumption, minimize performance loss, while meeting power budget | DVFS, VM consolidation, server power switching |
| Power and Performance Management of Virtualized Computing Environments via Lookahead Control, Kusic et al. [58] | Yes | CPU | Minimize power consumption, minimize performance loss | DVFS, VM consolidation, server power switching |
| Resource Allocation using Virtual Clusters, Stillwell et al. [59] | Yes | CPU | Maximize resource utilization, satisfy performance requirements | Resource throttling |
| Multi-Tiered On-Demand Resource Scheduling for VM-Based Data Center, Song et al. [60] | Yes | CPU, memory | Maximize resource utilization, satisfy performance requirements | Resource throttling |
| Shares and Utilities based Power Consolidation in Virtualized Server Environments, Cardosa et | Yes | CPU | Minimize power consumption, minimize performance loss | DFVS, soft scaling |

| Project name | Virtua-lization | System resources | Goal | Power-saving techniques |
|---|---|---|---|---|
| al. [61] | | | | |
| pMapper: Power and Migration Cost Aware Application Placement in Virtualized Systems, Verma et al. [62] | Yes | CPU | Minimize power consumption, minimize performance loss | DVFS, VM consolidation, server power switching |
| Resource pool management: Reactive versus proactive, Gmach et al. [63] | Yes | CPU, memory | Maximize resource utilization, satisfy performance requirements | VM consolidation, server power switching |
| GreenCloud: Energy-Efficient and SLA-based Management of Cloud Resources, Buyya et al. [64] [65] | Yes | CPU | Minimize energy consumption, satisfy performance requirements | Leveraging heterogeneity of Cloud data centers, DVFS |

# 8.1 Implications of Cloud Computing

Cloud computing has become a very promising paradigm for both consumers and providers in various areas including science, engineering and not to mention business. A Cloud typically consists of multiple resources possibly distributed and heterogeneous. Although the notion of a Cloud has existed in one form or another for some time now (its roots can be traced back to the mainframe era [66]), recent advances in virtualization technologies and the business trend of reducing the TCO in particular have made it much more appealing compared to when it was first introduced. There are many benefits from the adoption and deployment of Clouds, such as scalability and reliability; however, Clouds in essence aim to deliver more economical solutions to both parties (consumers and providers). By economical we mean that consumers only need to pay per their use and providers can capitalize poorly utilized resources. From the provider's perspective, the maximization of their profit is a high priority. In this regard, the minimization of energy consumption plays a crucial role. Recursively, energy consumption can be much reduced by increasing the resource utilization. Moreover, Cloud applications require movement of large data sets between the infrastructure and consumers, thus it is essential to consider both compute and network aspects of energy efficiency [67]. Energy usage in large-scale computing systems like Clouds also yields many other concerns including carbon emissions and system reliability.

Reduction in energy consumption by more effectively dealing with resource provisioning (avoidance of resource under/over provisioning) may be obtained [68]. Large profit-driven Cloud service providers typically develop and implement better power management, since they are interested in taking all necessary means to reduce energy costs to maximize their profit.

# 8.2 Non-Virtualized Systems

### 8.2.1 Load Management for Power and Performance in Clusters

Pinheiro et al. [21] have proposed a technique for managing a cluster of physical machines with the objective of minimizing the power consumption, while providing the required QoS. The authors claim that they present a new direction of research as all previous works deal with power efficiency in mobile systems or load balancing in clusters. The main technique to minimize power consumption is the load concentration, or unbalancing, while switching idle computing nodes off. The approach requires dealing with the power-performance trade-off, as performance of applications can be degraded due to the workload consolidation. The authors use the throughput and execution time of applications as constraints for ensuring the QoS. The nodes are assumed to be homogeneous. The algorithm periodically monitors the load and decides which nodes should be turned on or off to minimize the power consumption by the system, while providing expected performance. To estimate the performance the authors apply a notion of demand for resources, where resources include CPU, disk and network interface. This notion is used to predict performance degradation and throughput due to workload migration based on historical data. However, the demand estimation is static – the prediction does not consider possible demand changes over time. Moreover, due to sharing of the resource by several applications, the estimation of the resource demand for each application can be complex when the total demand exceeds 100% of the available resource capacity. For this reason, throughput degradation is not estimated in the experimental study. To determine the time to add or remove a node the authors introduce a total demand threshold that is set statically for each resource. This threshold is also supposed to solve the problem of the latency caused by a node addition, but may lead to performance degradation in the case of fast demand growth.

The actual load balancing is not handled by the system and has to be managed by the applications. The algorithm is executed on a master node that creates a single point of failure and might become a performance bottleneck in a large system. In addition, it is claimed that reconfiguration operations are time-consuming and the implementation of the algorithm adds or removes only one node at a time that may also be a reason for slow reaction in large-scale environments.

The authors have also investigated the cooperation between applications and OS in terms of power management decisions. They found that it can help to achieve more efficient control. However, the requirement for such cooperation leads to loss of the approach generality. Generality is also reduced as the system has to be configured for each application. This problem can be eliminated by application of virtualization technology. To evaluate the approach, the authors have conducted several experimental studies with different types of workloads: web-applications and compute intensive applications. The approach can be applied to multi-service mixed-workload environments with fixed SLA.

## 8.2.2 Managing Energy and Server Resources in Hosting Centers

Chase et al. [52] have studied the problem of managing resources in Internet hosting centers. Resources are shared among multiple service applications with specified SLA – the throughput and latency. The authors have developed an OS for an Internet hosting center (Muse) that is a supplement for operating systems of individual servers and supposed to manage and coordinate interactions between the data center's components. The main distinction from previous work is that the objective is not just to schedule resources efficiently, but also minimize the consumption of electrical power by the system components. In this work this approach is applied to data centers in order to reduce: operating costs (power consumption by computing resources and cooling system); carbon dioxide emissions, and thus the impact on the environment; thermal vulnerability of the system due to cooling failures or high service load; and over-provisioning in capacity planning. Muse addresses these problems by automatically scaling back the power demand (and therefore waste heat) when appropriate. Such a control over resource usage optimizes the trade-off between service quality and price, allowing the support of flexible SLA negotiated between consumers and a resource provider.

The main challenge is to determine resource demand of each application at its current request load level, and to allocate resources in the most efficient way. To deal with this problem the authors apply an economic framework: the system allocates resources in a way that maximizes the "profit" by balancing the cost of each resource unit against the estimated utility, or the "revenue" that is gained from allocating that resource unit to a service. Services "bid" for the resources in terms of volume and quality. This enables negotiation of the SLA according to the available budget and current QoS requirements, i.e. balancing cost of resource usage (energy cost) and benefit gained due to usage of this resource. This enables a data center to improve the energy efficiency under fluctuating workload, dynamically match load and power consumption, and respond gracefully to resource shortages.

The system maintains an active set of servers selected to serve requests for each service. Network switches are dynamically reconfigured to change the active set when necessary. Energy consumption is reduces by switching idle servers to power saving states (e.g. sleep, hibernation). The system is targeted at the web workload, which leads to "noise" in the load data. The authors address this problem by applying of the statistical "flip-flop" filter, which reduces the number of unproductive reallocations and leads to more stable and efficient control.

This work has created a foundation for the numereous studies in power efficient resource allocation at the data center level, however, the proposed approach has several weaknesses. The system deals only with CPU management, but does not take into account other system resources, such as memory, disk storage and network interface. It utilizes Advanced Power Management (APM), which is an outdated standard for Intel-based systems, while currently adopted by industry standard is ACPI. The thermal factor is not considered as well as the latency due to switching physical nodes on / off. The authors have pointed out that the management algorithm is stable, but it turns out to be relatively expensive during significant changes in th e workload. Moreover, heterogeneity of the software configuration requirements is not handled, which can be addressed by applying the virtualization technology.

### 8.2.3 Energy-Efficient Server Clusters

Elnozahy et al. [20] have explored the problem of power-efficient resource management in a single-service environment for web-applications with fixed SLA (response time) and auto load-balancing running on a homogeneous cluster. The motivation for the work is the reduction of operating costs and improvement of the error-proneness due to overheating. Two power management mechanisms are applied: switching physical nodes on and off (vary on vary off, VOVO) and DVFS of the CPU, whereas other system resources are not considered as they "consume a smaller fraction of the total system power consumption".

The authors have proposed five policies for resource management: Independent Voltage Scaling (IVS), Coordinated Voltage Scaling (CVS), Vary-On Vary-Off (VOVO), Combined Policy (VOVO-IVS) and Coordinated Combined Policy (VOVO-CVS). The last mentioned policy is stated to be the most advanced and is provided with a detailed description and mathematical model for determining CPU frequency thresholds. The thresholds define when it is appropriate to turn on an additional physical node or turn off an idle node. The main idea of the policy is to estimate total CPU frequency required to provide expected response time, determine the optimal number of physical nodes and set the proportional frequency on all the nodes.

The experimental results show that the proposed IVS policy can provide up to 29% energy savings and is competitive with more complex schemes for some workloads. VOVO policy can produce saving up to 42%, whereas coordinated voltage scaling policy in conjunction with VOVO (VOVO-CVS) results in 18% higher savings that are obtained using VOVO separately. However, the proposed approach is limited in the following factors. The transition time for starting up an additional node is not considered. Only a single application is assumed to be run on the cluster and the load-balancing is supposed to be done by an external system. Moreover, the algorithm is centralized that creates a single point of failure and reduces the system scalability. The workload data is not approximated, which can lead to inefficient decisions due to fluctuations in the demand. No other system resources except for CPU are considered in resource management decisions.

## 8.2.4 Energy-Aware Consolidation for Cloud Computing

Srikantaiah et al. [53] have investigated the problem of dynamic consolidation of applications serving small stateless requests in data centers to minimize the energy consumption. First of all, the authors have explored the impact of the workload consolidation on the energy-per-transaction metric depending on both CPU and disk utilizations. The obtained experimental results show that the consolidation influences the relationship between energy consumption and utilization of resources in a non-trivial manner. The authors have found that the energy consumption per transaction results in "U"-shaped curve. When the utilization is low, due to high fraction of the idle state, the resource is not efficiently used leading to a more expensive in terms of the energy-performance metric. On the other hand, high resource utilization results in increased cache miss rate, context switches and scheduling conflicts. Therefore, the energy consumption becomes high due to the performance degradation and consequently longer execution time. For the described experimental setup the optimal points of utilization are at 70% and 50% for CPU and disk utilizations respectively.

According to the obtained results, the authors stated that the goal of the energy-aware consolidation is to keep servers well utilized, while avoiding the performance degradation due to high utilization. They modeled the problem as a multi-dimensional bin packing problem, in which servers are represented by bins with each resource (CPU, disk, memory and network) considered as a dimension of the bin. The bin size along each dimension is defined by the determined optimal utilization level. The applications with known resource utilizations are represented by objects with an appropriate size in each dimension. The minimization of the number of bins is stated as leading to the minimization of the energy consumption due to switching off idle nodes. However, the model does not describe performance of applications that can be degraded due to the consolidation. Moreover, the energy consumption may depend on a particular set of application combined on a computer node.

The authors have proposed a heuristic for the defined bin packing problem. The heuristic is based on idea of minimization of the sum of the Euclidean distances of the current allocations to the optimal point at each server. As a request to execute a new application is received, the application is allocated to a server using the proposed heuristic. If the capacity of active servers is fulfilled, a new server is switched on, and all the applications are reallocated using the same heuristic in an arbitrary order. According to the experimental results, the energy used by the proposed heuristic is about 5.4% higher than optimal. The proposed approached is suitable for heterogeneous environments, however, it has several shortcomings. First of all, resource requirements of applications are assumed to be known a priory and constant. Moreover, migration of state-full applications between nodes incurs performance and energy overhead, which are not considered by the authors. Switching servers on / off also leads to significant costs that must be considered for a real-world system. Another problem with the approach is the requirement of an experimental study to obtain optimal points of the resource utilizations for each server. Furthermore, the decision about keeping the upper threshold of the resource utilization at the optimal point is not justified as the utilization above the threshold can symmetrically provide the same energy-per-transaction level.

## 8.2.5 Optimal Power Allocation in Server Farms

Gandhi et al. [54] have studied the problem of allocating an available power budget among servers in a virtualized heterogeneous server farm to minimize mean response time for HPC applications. The authors have investigated how server's CPU frequency scaling techniques affect the server's power consumption. They have conducted experiments applying DFS (T-states), DVFS (P-states) and DVFS+DFS (coarse grained P-states combined with fine grained T-states) for CPU intensive workloads. The results show linear power-to-frequency relationship for DFS and DVFS techniques and cubic square relationship for DVFS+DFS.

Given the power-to-frequency relationship, the authors consider the problem of finding the optimal power allocation as a problem of determining the optimal frequencies of the server's CPUs with ensuring minimization of the mean response time. To investigate the effect of different factors on the mean response time the authors have introduces a queuing theoretic model that allows prediction of the mean response time as a function of the power-to-frequency relationship, arrival rate, peak power budget, etc. The model also allows determining the optimal power allocation for every possible configuration of the above factors.

The approach has been experimentally evaluated against different types of workloads. The results show that an efficient power allocation can significantly vary for different workloads. To gain the best performance constrained by a power budget, it is not always optimal to run a small number of servers at their maximum speed. Oppositely, depending on the workload it can be more efficient to run more servers but at lower performance levels. The experimental results show that efficient power allocation can substantially improve server farm performance – up to a factor of 5 and by a factor of 1.4 on average.

### 8.2.6 Environment-Conscious Scheduling of HPC Applications

Garg et al. [55] have investigated the problem of energy and $CO_2$ efficient scheduling of HPC applications in geographically distributed Cloud data centers. The aim is to provide HPC users with the ability to leverage high-end computing resources supplied by Cloud computing environments on demand and in a pay-as-you-go basis. The authors have addressed the problem in the context of a Cloud resource provider and presented heuristics for energy-efficient meta-scheduling of applications across heterogeneous resource sites. Apart from reducing the maintenance costs, which results in higher profit for a resource provider, the proposed approach decreases carbon dioxide footprints. The proposed scheduling algorithms take into account energy cost, carbon emission rate, workload and CPU power efficiency, which change across different data centers depending on their location, design and resource management system.

The authors have proposed five scheduling policies, two of which minimize carbon dioxide emissions, two maximize the profit of resource providers, and the last one is a multi-objective policy that minimizes $CO_2$ emissions and maximizes the profit. The multi-objective policy finds for each application a data center, which provides the least carbon dioxide emissions, among data centers able to complete an application by its deadline. Then among all the application-data center pairs, the policy chooses one, which results in the maximal profit. These steps are repeated until all the applications are scheduled. The energy consumption is also reduced by applying DVFS for all the CPUs in data centers.

The proposed heuristics have been evaluated using simulations of different scenarios. The experimental results have shown that the energy-centric policies allow the reduction of energy costs by 33% on average. The proposed multi-objective algorithm can be effectively applied when limitations of carbon dioxide emissions are desirable by resource providers or forced by governments. This algorithm leads to reduced carbon emission rate, while maintains a high level of the profit.

## 8.3 Virtualized Systems

### 8.3.1 VirtualPower: Coordinated Power Management

Nathuji and Schwan [56] have investigated the problem of power efficient resource management in large-scale virtualized data centers. This is the first time when power management techniques have been explored in the context of virtualized systems. The authors have pointed out the following benefits of virtualization: improved fault and performance isolation between applications sharing the same resource; ability to relatively easy move VMs between physical hosts applying live or offline migration; support for hardware and software heterogeneity, which they investigated in their previous work [69]. Besides the hardware scaling and VMs consolidation, the authors apply a new power management technique in the context of virtualized systems called "soft resource scaling". The idea is to emulate hardware scaling by providing a VM less time for utilizing the resource using the VMM's scheduling capability. "Soft" scaling is useful when hardware scaling is not supported or provides a very small power benefit. The authors have found that combination of "hard" and "soft" scaling may provide higher power savings due to usually limited number of hardware scaling states.

The goals of the proposed approach are support for isolated and independent operation of guest VMs, and control and coordination of diverse power management policies applied by the VMs to resources. The system intercepts guest VMs' ACPI calls to perform changes in power states, map them on 'soft' states and uses as hints for actual changes in the hardware's power state. In this way the system supports guest VM's system level or application level specific power management policies, while maintaining isolation between multiple VMs sharing the same physical node.

The authors propose to split resource management into local and global policies. At the local level the system coordinates and leverages power management policies of guest VMs at each physical machine. An example of such a policy is the on-demand governor integrated into the Linux kernel. At this level the QoS is maintained as decisions about changes in power states are issued externally, by guest OS specific policies. However, the drawback of such a solution is that the power management may be inefficient due to a legacy or non power-aware guest OS. Moreover, power management decisions are usually done with some slack and the aggregated slack will grow with the number of VMs leading to under-optimal management. The authors have described several local policies aimed at the minimization of power consumption under QoS constraints, and at power capping. The global policies are responsible for managing multiple physical machines and use knowledge of rack- or blade-level characteristics and requirements. These policies consolidate VMs using migration in order to offload resources and place them into power saving states. The experiments conducted by the authors show that usage of the proposed system leads to efficient coordination of VM- and application-specific power management policies, and reduces the power consumption up to 34% with little or no performance penalties. However, the authors do not provide a detailed description of the global policies used limiting the analysis of the approach.

**8.3.2 Coordinated Multi-level Power Management**

Raghavendra et al. [57] have investigated the problem of power management for a data center environment by combining and coordinating five diverse power management policies. The authors argue that although a centralized solution can be implemented to handle all aspects of power management, it is more likely for a business environment that different solutions from multiple vendors will be applied. In this case it is necessary to solve the problem of coordination between individual controllers to provide correct, stable and efficient control. The authors classify existing solutions by a number of characteristics including the objective function, performance constraints, hardware / software and local / global types of policies. The range of solutions that fall into this taxonomy can be very wide. Therefore, instead of trying to address the wholes space, the authors focus on five individual solutions and propose five appropriate power management controllers. They have explored the problem in terms of control theory and apply feedback control loop to coordinate the controllers' actions.

The efficiency controller optimizes average power consumption by individual servers. The controller monitors the utilization of resources, based on the past history predicts future demand and appropriately adjusts the P-state of the CPU. The server manager implements power capping at the server level. It monitors power consumption by a server and reduces the P-state if the power budget is violated. The enclosure manager and the group manager implement power capping at the enclosure and data center level respectively. They monitor individual power consumptions across a collection of machines and dynamically re-provision power across systems to maintain the group power budget. The power budgets can be provided by system designers based on thermal or power delivery constraints, or by high level power managers. The VM controller reduces power consumption across multiple physical nodes by dynamically consolidating VMs and switching idle servers off. The authors provide integer programming model for the problem of optimization of VM allocation. However, the proposed model does not provide a protection from unproductive migrations due to workload fluctuations and does not show how SLA can be guaranteed in cases of fast changes in the workload. Furthermore, the transition time for reactivating servers as well as the ability to handle multiple system resources apart from the CPU are not considered.

The authors have provided experimental results, which show the ability of the system to reduce the power consumption under different workloads. The authors have pointed out an interesting outcome of the experiment: the actual power savings can vary depending on the workload, but "the benefits from coordination are qualitatively similar for all classes of workloads". In summary, the authors have presented the system for coordination of different power management policies. However, the proposed system is not able to ensure meeting QoS requirements as well as variable SLA from different applications. Therefore, the solution is suitable for enterprise environments, but not for Cloud computing providers, where more reliable QoS and a comprehensive support for SLA are essential.

### 8.3.3 Power and Performance Management via Lookahead Control

Kusic et al. [58] have explored the problem of power and performance efficient resource management in virtualized computing systems. The problem is narrowed to dynamic provisioning of VMs for multi-tiered web-applications according to current workload (number of incoming requests). SLA for each application are defined as the request processing rate. The clients pay for the provided service and receive refund in case of violated SLA as a penalty to the resource provider. The objective is to maximize resource provider's profit by minimizing both power consumption and SLA violation. The problem is stated as a sequential optimization and addressed using Limited Lookahead Control (LLC). Decision variables to be optimized are the following: the number of VMs to provision to each service; the CPU share allocated to each VM; the number of servers to switch on or off; and a fraction of incoming workload to distribute across the servers hosting the service.

The workload is assumed to be quickly changing, which means that resource allocations must be adapted over short time periods – "in order of tens seconds to a few minutes". Such requirement makes essential high performance of the optimization controller. The authors also incorporated in the model time delays and incurred costs for switching hosts and VMs on / off. Switching hosts on / off as well as resizing and dynamic consolidation of VMs via offline migration are applied as power saving mechanisms. However, DVFS is not performed due to low power reduction effect as argued by the authors.

The authors have applied Kalman filter to estimate the number of future requests, which is used to predict future system state and perform necessary reallocations. The authors have provided a mathematical model for the optimization problem. The utility function is risk-aware and includes risks of "excessive switching caused by workload variability" as well as transient power-consumption and opportunity costs. However, the proposed model requires simulation-based learning for the application specific adjustments: processing rate of VMs with different CPU shares must be known a priori for each application. This fact limits generality of the approach. Moreover, due to complexity of the model the optimization controller execution time reaches 30 minutes even for a small experimental setup (15 hosts), which is not suitable for large-scale real-world systems. The authors have applied neural networks to improve the performance; however, the provided experimental results are only for 10 hosts, and thus are not enough to prove the applicability of such a technique. The experimental results show that a server cluster managed using LLC saves 26% in the power consumption costs over a 24 hour period when compared to an uncontrolled system. Power savings are achieved with 1.6% SLA violations of the total number of requests.

## 8.3.4 Resource Allocation using Virtual Clusters

Stillwell et al. [59] have studied the problem of resource allocation for HPC applications in virtualized homogeneous clusters. The objective is to maximize resource utilization, while optimizing user-centric metric that encompasses both performance and fairness, which is referred to as the yield. The idea is to design a scheduler focusing on a user-centric metric. The yield of a job is "a fraction of its maximum achievable compute rate that is achieved". A yield of 1 means that the job consumes computational resources at its peak rate.

To formally define the basic resource allocation problem, the authors have assumed that an application requires only one VM instance; the application's computational power and memory requirements are static and known a priori. The authors have defined a Mixed Integer Programming Model that describes the problem. However, the solution of the model requires an exponential time, and thus can be obtained only for small instances of the problem. The authors have proposed several heuristics to solve the problem and evaluated them experimentally across different workloads. The results show that that the multi-capacity bin packing algorithm that sorts tasks in descending order by their largest resource requirement outperforms or equals to all the other evaluated algorithms in terms of minimum and average yield, as well as failure rate.

Subsequently, the authors have relaxed the stated assumptions and considered the cases of parallel applications and dynamic workloads. The researchers have defined a Mixed Integer Programming Model for the first case and adapted the previously designed heuristics to fit into the model. The second case allows migration of applications to address the variability of the workload, but the cost of migration is simplified and considered as a number of bytes required to transfer over network. To limit the overhead due to VM migration, the authors fix the amount of bytes that can be reallocated at one time. The authors have provided a Mixed Integer Programming Model for the defined problem; however, no heuristics have been proposed to solve large-scale problem instances. Limitations of the proposed approach are that no other system resources except for CPU are considered in the optimization and that the applications' resource needs are assumed to be known a priori, which is not typical in practice.

## 8.3.5 Multi-Tiered On-Demand Resource Scheduling

Song et al. [60] have studied the problem of efficient resource allocation in multi-application virtualized data centers. The objective is to improve the utilization of resource leading to reduced energy consumption. To ensure the QoS, the resources are allocated to applications proportionally according to the applications' priorities. Each application can be deployed using several VMs instantiated on different physical nodes. In resource management decisions only CPU and RAM utilizations are taken into account. In cases of limited resources, the performance of a low-priority application is intentionally degraded and the resources are allocated to critical applications. The authors have proposed scheduling at three levels: the application-level scheduler dispatches requests among application's VMs; the local-level scheduler allocates resources to VMs running on a physical node according to their priorities; the global-level scheduler controls the resource "flow" among applications. Rather than apply VM migration to implement global resource "flow", the authors pre-instantiate VMs on a group of physical nodes and allocate fractions of total amount of resources assigned to an application to different VMs.

The authors have presented a linear programming model for the resource allocation problem and heuristic for this model. They have provided the experimental results for three different applications running on a cluster: a web-application, a database and a virtualized office application showing that the approach allows satisfaction of the defined SLA. One of the limitations of the proposed approach is that it requires machine-learning to obtain the utility functions for applications. Moreover, it does not utilize VM migration to adapt the allocation in run-time. The approach is suitable for enterprise environments, where application can have explicitly defined priorities.

## 8.3.6 Shares and Utilities based Power Consolidation

Cardosa et al. [61] have investigated the problem of power-efficient VM allocation in virtualized enterprise computing environments. They leverage min, max and shares parameters, which are supported by the most modern VM managers. Min and max allow the user to specify minimum and maximum of CPU time that can be allocated to a VM. Shares parameter determines proportions, in which CPU time will be allocated to VMs sharing the same resource. Such approach suits only enterprise environments, as it does not support strict SLA and requires the knowledge of the applications' priorities.

The authors provide a mathematical formulation of the optimization problem. The objective function to be optimized includes the power consumption and utility gained from execution of a VM, which is assumed to be known a priori. The authors provide several heuristics for the defined model and experimental results. A basic strategy is to pack all the VMs at their maximum resource requirements in a first-fit manner and leave 10% of a spare capacity to handle the future growth of the resource usage. The algorithm leverages heterogeneity of the infrastructure by sorting physical machines in increasing order of the power cost per unit of capacity. The limitations of the basic strategy are that it does not leverage relative values of different VMs, it always allocates a VM at its maximum resource requirements and uses only 90% of each server's capacity. This algorithm has been used as the benchmark policy and improved throughout the paper eventually culminating in the recommended PowerExpandMinMax algorithm. In comparison to the basic policy, this algorithm uses the value of profit that can be gained by allocating an amount of resource to a particular VM. It leverages the ability to shrink a VM to min resource requirements when necessary and expand it when it is allowed by the spare capacity and can bring additional profit. The power consumption cost incurred by each physical server is deducted from the profit to limit the number of servers in use.

The authors have evaluated the proposed algorithms on a range of large scale simulations and a small real data center testbed. The experimental results show that the PowerExpandMinMax algorithm consistently outperforms the other policies across a broad spectrum of inputs – varying VM sizes and utilities, varying server capacities and varying power costs. One of the experiments on a real testbed showed that the overall utility of the data center can be improved by 47%. A limitation of this work is that migration of VMs is not applied in order to adapt the allocation of VMs in run-time – the allocation is static. Another problem is that no other system resources except for CPU are handled by the model. Moreover, the approach requires static definition of the applications' priorities that limits generality and applicability in real-world environments.

## 8.3.7 pMapper: Power and Migration Cost Aware Application Placement

Verma et al. [62] have investigated the problem of dynamic placement of applications in virtualized systems, while minimizing the power consumption and maintaining the SLA. To address the problem the authors have proposed the pMapper application placement framework. It consists of three managers and an arbitrator, which coordinates their actions and makes allocation decisions. Performance Manager monitors the applications' behavior and resizes VMs according to current resource requirements and the SLA. Power Manager is in charge of adjusting hardware power states and applying DVFS. Migration Manager issues instructions for live migration of VMs in order to consolidate the workload. Arbitrator has a global view of the system and makes decisions about new placements of VMs and determines which VMs and on which nodes should be migrated to achieve this placement. The authors claim that the proposed framework is general enough to be able to incorporate different power and performance management strategies under SLA constraints.

The authors have formulated the problem as a continuous optimization: at each time frame the VM placement should be optimized to minimize the power consumption and maximize the performance. They make several assumptions to solve the problem, which are justified by experimental studies. The first of them is the performance isolation, which means that a VM can be seen by an application running on that VM as a dedicated physical server with the characteristics equal to the VM parameters. The second assumption is that the duration of a VM live migration does not depend on the background load, and the cost of migration can be estimated a priori based on the VM size and profit decrease caused by an SLA violation. Moreover, the solution does not focus on specific applications and can be applied to any kind of the workload. Another assumption is that the power minimization algorithm can minimize the power consumption without knowing the actual amount of power consumed by the application.

The authors have presented several algorithms to solve the defined problem. They consider it as a bin packing problem with variable bin sizes and costs. The bins, items to pack and bin costs represent servers, VMs and power consumption of servers respectively. To solve the bin packing problem First-Fit Decreasing algorithm (FFD) has been adapted to work for differently sized bins with item-dependent cost functions. The problem has been divided into two sub-problems: in the first part, new utilization values are determined for each server based on the cost functions and required performance; in the second part, the applications are packed into servers to fill the target utilization. This algorithm is called min Power Packing (mPP). The first phase of mPP solves the cost minimization problem, whereas the second phase solves the application placement problem. mPP is also adapted to reduce the migration cost by keeping track of the previous placement while solving the second phase. This variant is termed mPPH. Finally, the placement algorithm has been designed that optimizes the power and migration cost trade-off (pMaP). A VM is chosen to be migrated only if the revenue due to the new placement exceeds the migration cost. pMap searches the space between the old and new placements and finds a placement that minimizes the overall cost (sum of the power and migration costs). The authors have implemented the pMapper architecture with the proposed algorithms and performed extensive experiments to validate the efficiency of the approach. The experimental results show that the approach allows saving about 25% of power relatively to the Static and Load Balanced Placement algorithms. The researchers suggest several directions for future work, such as consideration of memory bandwidth, more advanced application of idle states and extension of the theoretical prove of the problem.

**8.3.8 Resource Pool Management: Reactive Versus Proactive**

Gmach et al. [63] have studied the problem of energy-efficient dynamic consolidation of VMs in enterprise environments. The authors have proposed a combination of a trace-based workload placement controller and a reactive migration controller. The trace-based workload placement controller collects data on resource usage by VMs instantiated in the data center and uses this historical information to optimize the allocation, while meeting the specified quality of service requirements. This controller performs multi-objective optimization by trying to find a new placement of VMs that will minimize the number of server needed to serve the workload, while limiting the number of VM migrations required to achieve the new placement. The bound on the number of migrations is supposed to be set by the system administrator depending on the acceptable VM migration overhead. The controller places VMs according to their peak resource usage over the period since the previous reallocation, which is set to 4 hours in the experimental study.

The reactive migration controller continuously monitors the resource utilization of physical nodes and detects when the servers are overloaded or underloaded. In contrast to the trace-based workload placement controller, it acts based on the real-time data on resource usage and adapts the allocation in a small scale (every minute). The objective of this controller is to rapidly respond to fluctuations in the workload. The controller is parameterized by two utilization thresholds that determine overload and underload conditions. An overloading occurs when the utilization of CPU or memory of a server exceeds a given threshold. On the other hand, an underloading occurs when the CPU or memory usage averaged over all the physical nodes falls below a specified threshold. The threshold values are statically set depending on the performance analysis and quality of service requirements.

The authors have proposed several policies based on different combinations of the described optimization controllers with different utilization thresholds. The simulation-driven evaluation using three-months of real-world workload traces for 138 SAP applications has shown that the best results can be achieved by applying both optimization controllers simultaneously rather than separately. The best policy invokes the workload placement controller every 4 hours and when the servers are detected to be lightly utilized. The migration controller is executed in parallel to tackle overloading and underloading of servers when they occur. This policy provides minimal CPU violation penalties and requires 10-20% more CPU capacity than the ideal case.

### 8.3.9 GreenCloud: Energy-Efficient and SLA-based Management Cloud Resources

Buyya et al. [64] have proposed the GreenCloud project aimed at development of energy-efficient provisioning of Cloud resources, while meeting QoS requirements defined in SLA established through a negotiation between providers and consumers. The project has explored the problem of power-aware allocation of VMs in Cloud data centers for application services based on user QoS requirements such as the deadline and budget constraints [65]. The authors have introduced a real-time virtual machine model. Under this model, a Cloud provider provisions VMs for requested real-time applications and ensures meeting the specified deadline constraints.

The problem is addressed at several levels. At the first level, a user submits a request to a resource broker for provisioning resources for an application consisting of a set of sub-tasks with specified CPU and deadline requirements. The broker translates the specified resource requirements into a request for provisioning VMs and submits the request to a number of Cloud data centers. The data centers return the price of provisioning VMs for the broker's request if the deadline requirement can be fulfilled. The broker chooses the data center that provides the lowest price of resource provisioning. The selected data center's VM provisioner allocates the requested VMs to the physical resources, followed by launching the user's applications. The authors have proposed three policies for scheduling real-time VMs in a data center using DVFS to reduce the energy consumption, while meeting deadline constraints and maximizing the acceptance rate of provisioning requests. The Lowest-DVS policy adjusts the CPU's P-state to the lowest level, ensuring that all the real-time VMs meet their deadlines. The $\delta$-Advanced-DVS policy over-scales the CPU speed up to $\delta$% to increase the acceptance rate. The Adaptive-DVS policy uses the M/M/1 queueing model to calculate the optimal CPU speed if the arrival rate and service time of real-time VMs can be estimated in advance.

The proposed approach has been evaluated via simulations using the CloudSim toolkit [70]. The simulations results have shown that $\delta$-Advanced-DVS shows the best performance in terms of profit per unit of the consumed power, as the CPU performance is automatically adjusted according to the system load. The performance of Adaptive-DVS is limited by the simplified queueing model.

# 9 Conclusions and Future Directions

In recent years, energy efficiency has emerged as one of the most important design requirements for modern computing systems, such as data centers and Clouds, as they continue to consume enormous amounts of electrical power. Apart from high operating costs incurred by computing resources, this leads to significant emissions of carbon dioxide into the environment. For example, currently IT infrastructures contribute about 2% of total $CO_2$ footprints. Unless energy-efficient techniques and algorithms to manage computing resources are developed, IT's contribution in the world's energy consumption and $CO_2$ emissions is expected to rapidly grow. This is obviously unacceptable in the age of climate change and global warming. In this chapter, we have studied and classified different ways to achieve power and energy efficiency in computing systems. The recent developments have been discussed and categorized over the hardware, operating system, virtualization and data center levels.

Efficient power management in computing systems is a well-known and extensively studied in the past problem. Intelligent management of resources may lead to significant reduction of the energy consumption by a system, while meeting the performance requirements. Relaxation of the performance constraints usually results in further decreased energy consumption. Efficient resource management is extremely important for servers and data centers comprising multiple computer nodes. In large-scale data centers the cost of energy consumed by computing nodes and supporting infrastructure (e.g. cooling systems, power supplies, PDU) can exceed the cost of the infrastructure itself in a few years. One of the most significant advancements that has facilitated further development in the area is the implementation of the DVFS capability by hardware vendors and subsequent introduction of ACPI. These technologies have enabled software control over the CPU's power consumption traded for the performance. Managing power from this level is straightforward: the utilization of CPU is monitored, and its clock frequency and supply voltage pair is continuously adjusted to match current performance requirements. The maturity of this technique can be illustrated by the fact that widely spread Linux OS includes it as a kernel module. In this work we have classified and surveyed various approaches to control power consumption by the system from the OS level applying different power saving techniques and abstractions. The virtualization technology has advanced the area by introduction of a very effective power saving technique: consolidation of the workload in VMs to the minimal number of physical nodes and subsequent switching idle nodes off. Besides the consolidation, leading virtualization vendors (i.e. Xen, VMware) similarly to Linux OS implement continuous DVFS.

The power management problem becomes more complicated when considered from the data center level. In this case the system is represented by a set of interconnected computing nodes that need to be managed as a single resource in order to minimize the energy consumption. Live and offline migrations of VMs offered by the virtualization technology have enabled the technique of dynamic consolidation of VMs according to current performance requirements. However, VM migration leads to time delays and performance overhead, requiring careful analysis and intelligent techniques to eliminate non-productive migrations that can occur due to the workload variation. We have classified and discussed a number of the proposed approaches to deal with the problem of energy-efficient resource management in virtualized and non-virtualized data centers. Common limitations of the most of the works are that no other system resource except for CPU are considered in the optimization; transition time for switching power states of the resource and VM migration overhead are not handled leading to performance degradation; VM migration is not applied to optimize the allocation in run-time. More generic solution suitable for a modern Cloud computing environment should comply with the following requirements:

- Virtualization of the infrastructure to support hardware and software heterogeneity and simplify the resource provisioning.
- Application of VM migration to continuously adapt the allocation and quickly respond to changes in the workload.
- Ability to handle multiple applications with different SLA owned by multiple users.
- Guaranteed meeting of the QoS requirements for each application.
- Support for different kind of applications, mixed workloads.
- Decentralization and high performance of the optimization algorithm to provide scalability and fault tolerance.
- Optimization considering multiple system resources, such as CPU, memory, disk storage and network interface.

For the future research work we propose the investigation of the following directions. First of all, due to the wide adoption of multi-core CPUs, it is important to develop energy-efficient resource management approaches that will leverage such architectures. Apart from the CPU and memory, another significant energy consumer in data center is the network interconnect infrastructure. Therefore, it is crucial to develop intelligent techniques to manage network resources efficiently. One of the ways to achieve this for virtualized data centers is to continuously optimize network topologies established between VMs, and thus reduce network communication overhead and load of network devices. Another direction for future work, which deals with low-level system design, is improvement of the power supplies efficiency, as well as development of hardware components that support performance scaling proportionally to power consumption. Reduction of the transition overhead caused by switching between different power states and VM migration overhead can greatly advance energy-efficient resource management and has to be also addressed by future research. Cloud federations comprising geographically distributed data centers have to be leveraged to improve the energy efficiency. Efficient workload distribution across geographically distributed data centers can enable the reallocation of the workload to a place where energy or cooling is cheaper (e.g. solar energy during daytime across different time zones, efficient cooling due to climate conditions). Other important directions are providing fine grained user's control over power consumption / $CO_2$ emissions in Cloud environments and support for flexible SLA negotiated between resource providers and users. Building on the strong foundation of prior works, new projects are starting to investigate advanced resource management and power saving techniques. Nevertheless, there are many open challenges that become even more prominent in the age of Cloud computing.

# Acknowledgements

We would like to thank Adam Wierman (California Institute of Technology), Kresimir Mihic (Stanford University) and Saurabh Kumar Garg (University of Melbourne) for their constructive comments and suggestions on improving the paper.

# Appendix A. Operating system level research works.

| Project name | Approach / algorithm | Application adaptation | System resources | Target systems | Goal | Power saving techniques | Workload | Implementation |
|---|---|---|---|---|---|---|---|---|
| The Ondemand Governor, Pallipadi and Starikovskiy [19] | The OS continuously monitors the CPU utilization and sets the frequency and voltage according to performance requirements | No | CPU | Arbitrary | Minimize power consumption, minimize performance loss | DVFS | Arbitrary | Part of Linux kernel |
| ECOsystem, Zeng et al. [37] [38] | The system determines overall amount of currentcy and distributes it between applications according to their priorities. Applications expend currency by utilizing the resources | Applications must cooperate with the OS using power-based API | CPU, memory, disk storage, network interface | Mobile systems | Achieve target battery lifetime | Resource throttling | Arbitrary | Modified Linux kernel (introduced a new kernel thread kenrgd) |
| Nemesis OS, Neugebauer and McAuley [39] | Nemesis notifies applications if their energy consumption exceeds the threshold. The applications must adapt their behaviour according to the signals from the OS | Applications must be able to adapt their behavior according to the signals from the OS | CPU, memory, disk storage, network interface | Mobile systems | Achieve target battery lifetime | Resource throttling | Real-time applications | New operating system, source codes are available to download |
| GRACE, Sachs et al. [40] [41] | Three levels of adaptation: global, per-application and internal. All the adaptation levels are coordinated to ensure adaptation effective across all levels | Applications must be able to adapt their behavior according to the signals from the OS | CPU, network interface | Mobile systems | Minimize energy consumption, satisfy performance requirements | DVFS, resource throttling | Real-time multimedia applications | Extension of Linux OS |
| Linux/RK, Rajkumar et al. [42] | Proposed four alternative DVFS algorithms. Each is suitable for different system characteristics and is selected automatically by the system | No | CPU | Real-time systems | Minimize energy consumption, satisfy performance requirements | DVFS | Arbitrary | Real-time extensions to the Linux kernel |
| Coda and Odyssey, Flinn and Satyanarayanan [43] | Coda implements application-transparent adaptation in the context of a distributed file system. Odyssey implements application adaptation allowing adjustment of the resource | Applications must be able to adapt their behavior according to the signals from the OS | CPU, network interface | Mobile systems | Minimize energy consumption allowing application data degradation | Resource throttling | Multimedia applications | Coda is implemented as a package for Linux, Odyssey is integrated into Linux |

| Project name | Approach / algorithm | Application adaptation | System resources | Target systems | Goal | Power saving techniques | Workload | Implementation |
|---|---|---|---|---|---|---|---|---|
| | consumption by the cost of output data quality | | | | | | | |
| PowerNap, Meisner et al. [44] | Leveraging short idle periods in the resource utilization using fast transitions to system-wide low power states | No | System-wide | Server systems | Minimize power consumption, minimize performance loss | DCD | Arbitrary | Extension to Linux OS |

# Appendix B. Data center level research works.

| Project name | Virtua-lization | Approach / algorithm | System resources | Target systems | Goal | Power saving techniques | Workload | Implementation |
|---|---|---|---|---|---|---|---|---|
| Load Balancing and Unbalancing for Power and Performance in Cluster-Based System, Pinheiro et al. [21] | No | The system periodically monitors the load and decides which nodes should be turned on or off to minimize power consumption by the system, while providing expected performance. | CPU, disk storage, network interface | Homogeneous | Minimize power consumption, minimize performance loss | Server power switching | Arbitrary | Extension of Linux |
| Managing energy and server resources in hosting centers, Chase et al. [52] | No | Economical framework: the system allocate resources in a way to maximize "profit" by balancing the cost of each resource unit against the estimated utility or "revenue" that is gained from allocating that resource unit to a service. Services "bid" for the resources in terms of volume and quality. The system maintains an active set of servers selected to serve requests for each service. Energy consumption is reduces by switching idle servers to power saving states. | CPU | Homogeneous | Minimize power consumption, minimize performance loss | Workload consolidation, server power switching | Web-applications | Extension of FreeBSD OS |
| Energy-Efficient Server Clusters, Elnozahy et al. [20] | No | The system estimates total CPU frequency required to provide expected response time, determine the optimal number of physical nodes and set the proportional frequency on all the nodes. The thresholds define when it is appropriate to turn on an additional physical node or turn off an idle node. | CPU | Homogeneous | Minimize energy consumption, satisfy performance requirements | DVFS, server power switching | Web-applications | Simulation |
| Energy-aware | No | Applications are allocated to servers using a | CPU, | Heterogeneous | Minimize | Workload | Online | Simulation |

| Project name | Virtua-lization | Approach / algorithm | System resources | Target systems | Goal | Power saving techniques | Workload | Implementation |
|---|---|---|---|---|---|---|---|---|
| Consolidation for Cloud Computing, Srikantaiah et al. [53] | | heuristic for multi-dimensional bin packing, resulting in the desired workload distribution across servers. If the request cannot be allocated, a new server is turned on and all requests are re-allocated using the same heuristic, in an arbitrary order. | disk storage | | energy consumption, satisfy performance requirements | consolidation, server power switching | services | |
| Optimal Power Allocation in Server Farms, Gandhi et al. [54] | No | Queueing theoretical model is used to predict the mean response time as a function of power-to-frequency relationship, arrival rate, peak power budget, etc. The model also allows determining the optimal power allocation for every possible configuration of the above factors. | CPU | Heterogeneous | Allocate the available power budget to minimize mean response time | DVFS | Web-applications | Simulation |
| Environment-Conscious Scheduling of HPC Applications, Garg et al. [55] | No | Five heuristics for scheduling HPC applications across geographically distributed Cloud data centers with the objective of minimization of energy consumption and carbon emissions, and maximization of the resource provider's profit. | CPU | Heterogeneous | Minimize energy consumption and $CO_2$ emissions, maximize profit | DVFS, leveraging geographical distribution of data centers | HPC applications | Simulation |
| VirtualPower: Coordinated Power Management in Virtualized Enterprise Systems, Nathuji and Schwan [56] | Yes | Hierarchical power management: at the local level the system coordinates and leverages power management policies of guest VMs at each physical machine; global policies are responsible for managing multiple physical machines and have knowledge about rack- or blade-level characteristics and requirements. | CPU | Heterogeneous | Minimize energy consumption, satisfy performance requirements | DFVS, soft scaling, VM consolidation, server power switching | Arbitrary | Extension of Xen |
| Coordinated Multi-level Power Management for the Data Center, Raghavendra et al. [57] | Yes | A combination of five individual power management solutions that are coordinatively act across a collection of machines and dynamically re-provision power across them to meet the power budget. | CPU | Heterogeneous | Minimize power consumption, minimize performance loss, meet power budget | DVFS, VM consolidation, server power switching | Arbitrary | Combining and cooperation of five independent commercial solutions |
| Power and Performance Management of Virtualized Computing Environments via | Yes | The behavior of each application is captured using simulation-based learning. A limited look-ahead control (LLC) is applied to estimate future system states over a prediction horizon using Kalman filter. | CPU | Heterogeneous | Minimize power consumption, minimize performance loss | DVFS, VM consolidation, server power switching | Online services | VMware API, Linux shell commands and IPMI |

| Project name | Virtua­lization | Approach / algorithm | System resources | Target systems | Goal | Power saving techniques | Workload | Implementation |
|---|---|---|---|---|---|---|---|---|
| Lookahead Control, Kusic et al. [58] | | | | | | | | |
| Resource Allocation using Virtual Clusters, Stillwell et al. [59] | Yes | The authors have proposed several heuristics to solve the resource allocation problem and evaluated them experimentally across different workloads. The results show that that the multi-capacity bin packing algorithm that sorts tasks in descending order by their largest resource requirement outperforms or equals to all the other evaluated algorithms in terms of minimum and average yield, as well as failure rate. | CPU | Homogeneous | Maximize resource utilization, satisfy performance requirements | Resource throttling | HPC applications | Extension of Xen |
| Multi-Tiered On-Demand Resource Scheduling for VM-Based Data Cente, Song et al. [60] | Yes | Three scheduling levels: the application-level scheduler dispatches requests among application's VMs; the local-level scheduler allocates resources to VMs running on a physical node according to their priorities; the global-level scheduler controls the resource "flow" among applications. | CPU, memory | Heterogeneous | Maximize resource utilization, satisfy performance requirements | Resource throttling | Arbitrary | Extension of Xen |
| Shares and Utilities based Power Consolidation in Virtualized Server Environments, Cardosa et al. [61] | Yes | The hypervisor distributes resources among VMs according to a sharing based mechanism, assuming that the minimum and maximum amounts of resources that can be allocated to a VM are specified. | CPU | Heterogeneous | Minimize power consumption, minimize performance loss | DFVS, soft scaling | Arbitrary | Extension of VMware ESX |
| pMapper: Power and Migration Cost Aware Application Placement in Virtualized Systems, Verma et al. [62] | Yes | The authors consider the problem as continuous optimization and address it using heuristics for the bin packing problem. Performance Manager monitors applications behavior and resize VMs according to current resource requirements and the SLA. Power Manager adjusts hardware power states and applies DVFS. Migration Manager issues instructions for live migration of VMs. Arbitrator makes decisions about new placements of VMs and determines VMs to migrate. | CPU | Heterogeneous | Minimize power consumption, minimize performance loss | DVFS, VM consolidation, server power switching | Arbitrary | Extension of VMware ESX |

| Project name | Virtua-lization | Approach / algorithm | System resources | Target systems | Goal | Power saving techniques | Workload | Implementation |
|---|---|---|---|---|---|---|---|---|
| Resource pool management: Reactive versus proactive, Gmach et al. [63] | Yes | The authors apply a combination of two optimization controllers: proactive global optimization using the workload placement controller and reactive adaptation using the migration controller. | CPU, memory | Heterogeneous | Maximize resource utilization, satisfy performance requirements | VM consolidation, server power switching | Arbitrary | Simulation |
| GreenCloud: Energy-Efficient and SLA-based Management of Cloud Resources, Buyya et al. [64], [65] | Yes | The project has proposed energy-efficient provisioning of Cloud resources along with meeting users' QoS requirements as defined in SLAs. The authors have developed heuristics for scheduling real-time VMs in Cloud data centers applying DVFS in order to minimize the energy consumption and deadline constraints of the applications. | CPU | Heterogeneous | Minimize energy consumption, satisfy performance requirements | Leveraging heterogeneity of Cloud data centers, DVFS | HPC applications | Simulation |

# About the Authors

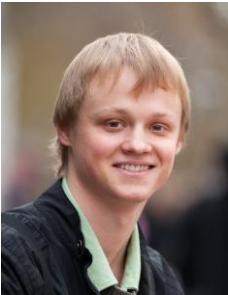

**Anton Beloglazov** is a PhD Candidate at the Cloud Computing and Distributed Systems (CLOUDS) Laboratory within the Department of Computer Science and Software Engineering at the University of Melbourne, Australia. He has completed his Bachelor's and Master's degrees in Informatics and Computer Science at the faculty of Automation and Computer Engineering of Novosibirsk State Technical University, Russian Federation. Under his PhD studies, Anton is actively involved in research on energy- and performance-efficient resource management in virtualized data centers for Cloud computing. He has been contributing to the development of the CloudSim toolkit, a modern open-source framework for modeling and simulation of Cloud computing infrastructures and services. Anton has publications in internationally recognized conferences and journals. He is a frequent reviewer for research conferences and journals.

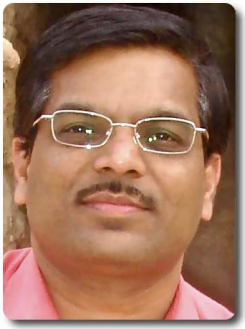

**Dr. Rajkumar Buyya** is Professor of Computer Science and Software Engineering; and Director of the Cloud Computing and Distributed Systems (CLOUDS) Laboratory at the University of Melbourne, Australia. He is also serving as the founding CEO of Manjrasoft Pty Ltd., a spin-off company of the University, commercializing its innovations in Grid and Cloud Computing. He has authored and published over 300 research papers and four text books. The books on emerging topics that Dr. Buyya edited include, High Performance Cluster Computing (Prentice Hall, USA, 1999), Content Delivery Networks (Springer, Germany, 2008), Market-Oriented Grid and Utility Computing (Wiley, USA, 2009), and Cloud Computing (Wiley, USA, 2019). He is one of the highly cited authors in computer science and software engineering worldwide. Software technologies for Grid and Cloud computing developed under Dr. Buyya's leadership have gained rapid acceptance and are in use at several academic institutions and commercial enterprises in 40 countries around the world. Dr. Buyya has led the establishment and development of key community activities, including serving as foundation Chair of the IEEE Technical Committee on Scalable Computing and four IEEE conferences (CCGrid, Cluster, Grid, and e-Science). He has presented over 250 invited talks on his vision on IT Futures and advanced computing technologies at international conferences and institutions in Asia, Australia, Europe, North America, and South America. These contributions and international research leadership of Dr. Buyya are recognized through the award of "2009 IEEE Medal for Excellence in Scalable Computing" from the IEEE Computer Society, USA. Manjrasoft's Aneka technology for Cloud Computing developed under his leadership has received "2010 Asia Pacific Frost & Sullivan New Product Innovation Award".

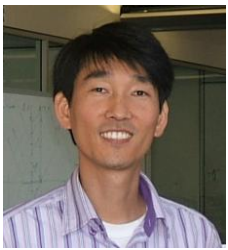

**Dr. Young Choon Lee** received the Ph.D. degree in problem-centric scheduling in heterogeneous computing systems from the University of Sydney in 2008. He received Best Paper Award from the 10th IEEE/ACM International Symposium on Cluster, Cloud and Grid Computing (CCGrid 2010). He is a member of the IEEE. His current research interests include scheduling strategies for heterogeneous computing systems, nature-inspired techniques, and parallel and distributed algorithms.

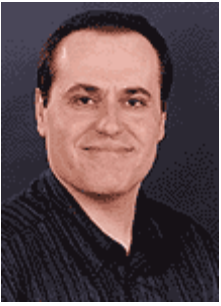**Dr. Albert Y. Zomaya** is currently the Chair Professor of High Performance Computing and Networking in the School of Information Technologies, The University of Sydney. He is the author/co-author of seven books, more than 350 publications in technical journals and conferences, and the editor of eight books and eight conference volumes. He is currently an associate editor for 16 journals, such as the IEEE Transactions on Computers, IEEE Transactions on Parallel and Distributed Systems and the Journal of Parallel and Distributed Computing. He is also the Founding Editor of the Wiley Book Series on Parallel and Distributed Computing and was the Chair the IEEE Technical Committee on Parallel Processing (1999-2003) and currently serves on its executive committee. He also serves on the Advisory Board of the IEEE Technical Committee on Scalable Computing and IEEE Systems, Man, and Cybernetics Society Technical Committee on Self-Organization and Cybernetics for Informatics and is a Scientific Council Member of the Institute for Computer Sciences, Social-Informatics, and Telecommunications Engineering (in Brussels). Professor Zomaya is also the recipient of the Meritorious Service Award (in 2000) and the Golden Core Recognition (in 2006), both from the IEEE Computer Society. He is a Chartered Engineer, a Fellow of the American Association for the Advancement of Science, the IEEE, the Institution of Engineering and Technology (U.K.), and a Distinguished Engineer of the ACM. His research interests are in the areas of distributed computing, parallel algorithms and mobile computing.